\documentclass[reprint,aps,prl,amssymb, amsmath,superscriptaddress]{revtex4-1}
\usepackage{graphicx}
\usepackage{pifont}
\usepackage{hyperref}
\usepackage{bm} 
\hypersetup{
    colorlinks=true, 
    linktoc=all,     
    linkcolor=blue,  
    citecolor =blue
}
\usepackage[T1]{fontenc}
\let\vaccent=\v 
\renewcommand{\v}[1]{\bm{#1}} 
\newcommand{\avg}[1]{\langle#1\rangle}
\newcommand{\abs}[1]{\left| #1 \right|} 

\newcommand{\dl}{\frac{d}{d\ell}}
\usepackage{etoolbox}

\begin{document}

\title{Novel criticality of Dirac fermions in the presence of emergent gauge fields}
\author{Elliot Christou}
\affiliation{London Centre for Nanotechnology, University College London, Gordon St., London, WC1H 0AH, United Kingdom}
\author{Fernado de Juan}
\affiliation{Donostia International Physics Center, 20018 Donostia-San Sebastian, Spain}
\affiliation{IKERBASQUE, Basque Foundation for Science, Maria Diaz de Haro 3, 48013 Bilbao, Spain}
\author{Frank Kr\"uger}
\affiliation{London Centre for Nanotechnology, University College London, Gordon St., London, WC1H 0AH, United Kingdom}
\affiliation{ISIS Facility, Rutherford Appleton Laboratory, Chilton, Didcot, Oxfordshire OX11 0QX, United Kingdom}
\begin{abstract}
We consider spontaneous symmetry breaking transitions of strongly interacting two-dimensional Dirac fermions minimally coupled to mass and emergent gauge field order parameters. Using a renormalization group analysis, we show that the presence of gauge fields leads to novel fermion induced quantum (multi-)criticality where the putative emergent Lorentz invariance is violated. We illustrate this with the example of translational symmetry breaking due to charge-density wave order on the honeycomb lattice. Finally, we identify that topological phase transitions are well described by this effective field theory.
\end{abstract}
\date{\today}

\maketitle

Interacting Dirac fermions exhibit the simplest form of fermionic quantum criticality.
In high energy physics this has been known for some time as spontaneous fermion mass generation and chiral symmetry breaking in the Gross-Neveu-Yukawa (GNY) model~\cite{grossneveuprd1974,zinnjustinnpb1991}.
The prototypical condensed matter examples are semimetal-insulator transitions on the half-filled honeycomb lattice~\cite{herbutprl2006,herbutetalprb2009}, which are driven by strong on-site and nearest-neighbor repulsive interactions.
The low-energy excitations are well described by Dirac fermions~\cite{semenoffprl1984}, which couple to the dynamical order parameter fields and play a crucial role in determining the universal behavior~\cite{zinnjustin,sachdev}.
In recent years GNY models and sign-free lattice Quantum Monte Carlo simulations have helped to push our understanding of fermionic criticality beyond the Landau-Ginzburg-Wilson paradigm~\cite{senthiletalscience2004,lietalnatcom2017,schererherbutprb2016,jianyaoprb2017a,satoetalprl2017,classenetalprb2017,torresetalprb2018,janssenetalprb2018,royjuricicprb2019,xuetalprx2019, lietalarxiv2019}. 
 
Strong parallels can be drawn between the high energy and condensed matter settings.
Firstly, the quantum critical fixed point exhibits emergent Lorentz invariance with a characteristic terminal velocity and dynamical exponent  $z=1$~\cite{vojtaetalprl2000,anberdonoghueprd2011,royetaljhep2016,janssenheprb2017,royetalprl2018,langlauchliprl2019}.
Secondly, the emergent chiral symmetry is spontaneously broken in the ordered phase, opening a mass gap in the Dirac spectrum. 

Yet it is precisely the reduction from Poincar\'{e} symmetry to the crystallographic space groups that allows the solid-state to host exotic fermionic quasiparticles 
with no elementary particle analogs~\cite{bradlynetalscience2016}  and interesting criticality. A prominent example are anisotropic 
semimetals~\cite{yangetalnatphys2014,surroyprl2019} displaying relativistic and non-relativistic dynamics along orthogonal directions.  
Such unconventional excitations  are known to exist at the critical point of topological phase transitions involving nodal fermions~\cite{yangetalprl2013}
and the underlying quasiparticles have been dubbed semi-Dirac fermions~\cite{dietletalprl2008,montambauxetalprb2009,banerjeeetalprl2009}. 
\begin{figure}[t]
 \includegraphics[width=0.4\textwidth]{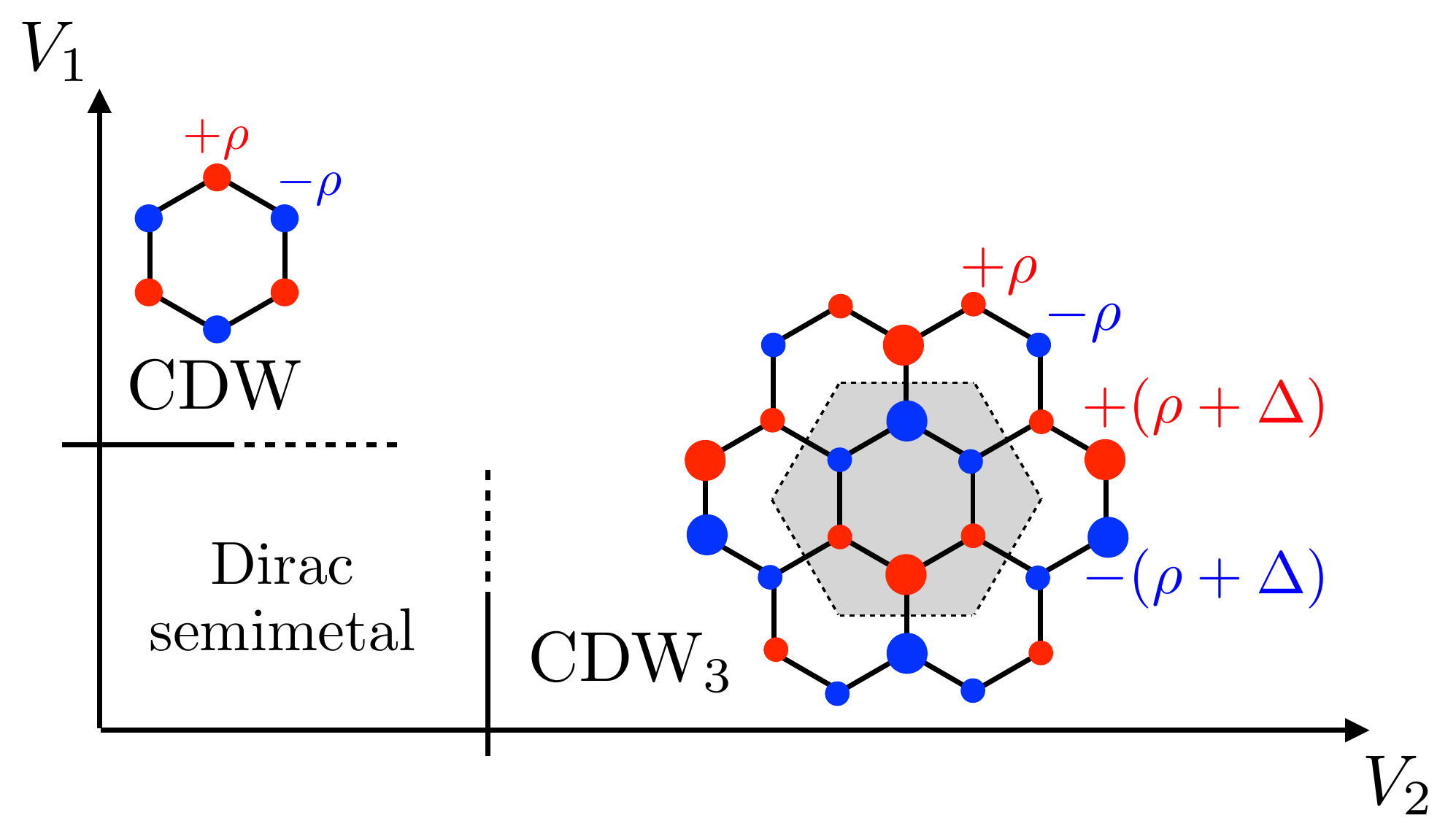}
 \caption{Schematic phase diagram of Dirac fermions on the half-filled
   honeycomb lattice with first and second neighbor repulsive
   interactions $V_1$ and $V_2$. Charge modulation is shown relative
   to half-filling.
 The CDW breaks sublattice inversion symmetry. The CDW$_3$ breaks 
 translation, rotation, mirror and inversion symmetries.}
\label{phasediagram}
\end{figure}

In this paper we investigate the universal nature of Dirac fermion systems undergoing quantum phase transitions with dynamical order parameter fields that minimally 
couple to the  fermions as components of \textit{emergent} gauge fields, in addition to the standard mass fields of GNY theories. We find that the quantum critical 
points violate the putative emergent Lorentz invariance of the GNY critical fixed points.

Although our conclusions are general, we illustrate these principles by means of a concrete example. We examine the half-filled two-dimensional honeycomb 
lattice in the presence of dominant next-nearest neighbor repulsions. Extensive numerical ~\cite{grushinetalprb2013,garciamartinezetalprb2013,daghoferhohenadlerprb2014,djuricetalprb2014,capponilauchliprb2015,motruketalprb2015,
schereretalprb2015,volpezetalprb2016,delapenaetalprb2017,kuritaetalprb2016,bijelicetalprb2018} and analytical~\cite{christouetalprb2018} efforts have concluded 
that fluctuations beyond mean-field theory lead to a direct transition from the Dirac semimetal to  charge density wave order (CDW$_3$) with a threefold increased 
unit cell (see Fig.~\ref{phasediagram}). 

From the lattice model we derive the effective field theory that describes the onset of CDW$_3$ order at half-filling.
The theory is formulated in terms of the irreducible representations of all possible charge instabilities of the six-site unit cell. While one representation couples like
the mass field in the GNY theory, the others, crucial for the lattice symmetry breaking, couple like non-Abelian gauge fields. This gauge structure is an emergent 
property of the continuum field theory that is broken by the lattice. Importantly, the lattice terms turn out to be irrelevant at the critical fixed point, which implies the scaling is universal. Naturally, this effective gauge symmetry is broken on either side of the transition.

We point out that the apparent local gauge structure that emerges at low energies in the continuum limit is physically distinct from the local lattice gauge structure present 
at all energy scales in, for example, spin liquids.

We characterize the fermion-induced (multi-)critical fixed point using the renormalization group (RG), which is controlled with the $\epsilon$ expansion and by generalizing to a large number $N$ of fermion flavors.
We find that the universality class is quantitatively distinct from the putative chiral GNY class, which highlights the relevant role of spontaneous lattice symmetry breaking.
Next, in combination with a free energy analysis, our results show that the broken symmetry state is in the vicinity of a topological critical point.
This demonstrates that topological phase transitions in Dirac systems can be comprehensively 
described by effective field theories containing mass fields and emergent non-Abelian gauge fields.

This paper is organized as follows: In Sec~\ref{seceffectivetheory}, we derive the low-energy theory of lattice symmetry breaking charge order on the honeycomb lattice. We also discuss some important symmetries from the lattice, as well as those that emerge in the low-energy, continuum limit. In Sec.~\ref{secrgequations} the RG equations for the effective field theory are presented, and their critical fixed-point solutions are studied. In Sec.~\ref{secbrokensymmetry} the broken symmetry state is analyzed by minimizing the free energy obtained from integrating out the fermions. In addition, a generic description of topological phase transitions in Dirac systems is presented. Conclusions are drawn in Sec.~\ref{secdiscussion}.

\section{Effective Theory} 
\label{seceffectivetheory}
We consider spin-polarized fermions on the honeycomb lattice at half-filling, subject to first and second neighbor repulsive interactions  $V_1$ and $V_2$, 
\begin{equation}
\hat{H} = -t \sum_{\avg{i,j}} (c_i^\dagger c_j + \text{h.c.}) + V_1\sum_{\avg{i,j}}\hat{n}_i \hat{n}_j +V_2\sum_{\avg{\avg{i,j}}}\hat{n}_i \hat{n}_j,
\label{lattice}
\end{equation}
where $t$ is the  hopping amplitude and $c_i$ and $\hat{n}_i=c^\dagger_i c_i$ denote the fermion annihilation and density operators.

\subsection{Non-Interacting Effective Theory}
In the absence of interactions, the  low-energy theory of the semimetallic state describes massless Dirac fermions  at the valleys $\v{K}_{\tau=\pm}=\frac{4\pi}{3\sqrt{3}}(\tau,0)$~\cite{castronetoetalrmp2009},
\begin{equation}
 H_0=
 v_F\int^\Lambda\frac{d^2\v{k}}{(2\pi a )^2}\,\Psi^{\dagger}(\v{k})\, \v{k}\cdot\v{\alpha}\,\Psi(\v{k}),\label{dirac}
\end{equation}
which is defined up to the ultraviolet cut-off $\Lambda\sim 1/a$ where $a$ is the lattice spacing.
Here $v_F=3ta/2$ is the Fermi velocity, $\v{k}=(k_x,k_y)$,  and $\v{\alpha}=(\alpha_x,\alpha_y)$.
From hereon we work in units where $a=1$. 
For compactness, we have introduced the four-component spinor
\begin{equation}
  \Psi = (\psi_{A+},\psi_{A-},\psi_{B+},\psi_{B-}),
  \label{diracsubspace}
\end{equation}
and tensor products
\begin{equation}
(  \alpha_x,\alpha_y,\alpha_z) =(\sigma_x\otimes \tau_z, \sigma_y\otimes\tau_0,\sigma_z\otimes\tau_z),
\end{equation} 
where  $\sigma_\mu,\, \tau_\mu$ ($\mu =0, x,y,z$)  are the four-vectors of identity and Pauli matrices acting respectively on the spatial honeycomb sublattice $\sigma=A,B$ and Dirac valley $\tau=\pm$ pseudospins.
The $SU(2)$ pseudospin algebra $[\alpha_i,\alpha_j]=2 i\sum_{k=x}^{z}\epsilon_{ijk}\alpha_k$ and  $\{\alpha_i,\alpha_j\}=2\delta_{ij}\alpha_0$ is formed, with the (implicit) identity  $\alpha_0=\sigma_0\otimes\tau_0$.

\subsubsection{Emergent Symmetries}
In addition to the symmetries of the honeycomb lattice, Eq.~\eqref{dirac} is endowed with a number of emergent symmetries.
These include: (i) The intravalley spatial rotational symmetry
\begin{equation}
  \label{rotationalsymmetry}
  \begin{gathered}
    \Psi \rightarrow e^{i \theta \alpha_z/2}\Psi,\\
    \v{k} \rightarrow (k_x\cos\theta -k_y\sin\theta,k_x\cos\theta+k_y\sin\theta).
\end{gathered} 
\end{equation}
(ii) The global $SU(2)$ chiral/gauge symmetry
\begin{equation}
  \Psi \rightarrow e^{i \sum_{i=1}^3 \theta^i T^i}\Psi,
  \label{chiralgaugesymmetry}
\end{equation}
that is generated by
\begin{equation}
  (T^1,T^2,T^3) = (-\sigma_y\otimes\tau_y,\sigma_y\otimes\tau_x,\sigma_0\otimes\tau_z).
\end{equation}
$T^{1,2,3}$ form a second $SU(2)$ algebra, with the (implicit) identity $T^0=\sigma_0\otimes
\tau_0$.
(iii) The pseudo-relativistic invariance where the ``speed of light'' is $v_F$.
This Lorentz invariance is exposed using the anticommuting Dirac matrices, in which the covariant form of the non-interacting local Lagrangian is achieved,
\begin{equation}
  L_\Psi =  \bar{\Psi}[\partial_\tau \gamma_0 + v_F \v{\partial}\cdot \v{\gamma}]\Psi,
  \label{psigamma}
\end{equation}
where  $\bar{\Psi}= \Psi^\dagger \gamma_0 $, $\gamma_0 = \alpha_z T^3$, $\v{\gamma} =i  \gamma_0 \v{\alpha}$, and 
\begin{equation}
\Psi(\tau,\v{r}) =\int^{\infty}_{-\infty} \frac{d k_0}{2\pi} \int^\Lambda \frac{d^2\v{k}}{(2\pi )^2} \,
e^{-i k_0\tau-i\v{k}\cdot\v{r}}\Psi(k_0,\v{k}).
\end{equation}
Here $\tau$ and $k_0$ are imaginary time and frequency, $\v{r}=(x,y)$ is position and $\v{\partial} = (\partial_x,\partial_y)$ is the spatial gradient.

\subsubsection{Symmetries from the Honeycomb Lattice}
Eq.~\eqref{dirac} inherits the symmetries of the honeycomb lattice point group $C_{6v}$, which includes reflections, rotations and inversions~\cite{baskoprb2008,herbutetalprb2009b,winklerzulickeanziam2015}.
The reflection symmetries in the $x$ and $y$ planes are
\begin{equation}
\begin{aligned}
R_x&: \Psi\rightarrow  \alpha_x T^3 \Psi,\,\,  k_y\rightarrow -k_y ,\\
R_y&: \Psi\rightarrow \alpha_yT^2\Psi,\,\, k_x\rightarrow -k_x.
\end{aligned}
\end{equation}
$R_x$ interchanges the sublattices, as can be seen from the lattice in Fig.~\ref{phasediagram}. $R_y$ interchanges the Dirac valleys.
Together $R_xR_y$ defines spatial inversion, which is equivalent to a $C_2$ rotation.
The honeycomb lattice also has threefold rotational symmetry 
\begin{equation}
\begin{gathered}C_3:\Psi \rightarrow e^{\pm   2\pi i \alpha_z/3}\Psi,\\ \v{k} \rightarrow (-k_x\mp\sqrt{3}k_y,\pm\sqrt{3}k_x-k_y)/2.
\end{gathered}
\end{equation}
There is also the translational symmetry under the primitive lattice vectors 
\begin{equation}
\v{a}_1=\frac{\sqrt{3}}{2}(
1,\sqrt{3}),\,\, \v{a}_2 =  \frac{\sqrt{3}}{2}(-
1,\sqrt{3}),
\end{equation}
that is generated by $T^3$ (also the generator of chiral transformations), and is enacted as
\begin{equation}
t_{\v{a}_{1,2}} : \Psi(\v{r}) \rightarrow e^{i \v{K}_+ \cdot \v{a}_{1,2} T^3} \Psi(\v{r}+\v{a}_{1,2}),\label{translations}
\end{equation}
where $\v{K}_+ \cdot \v{a}_{1,2} = \pm2\pi/3$.
In the following, we are interested in the case where this primitive translational symmetry, in combination with point group symmetries, is spontaneously broken.
We refer to this as spontaneous lattice symmetry breaking.

Additionally, the theory has the discrete time reversal $\mathcal{T}$, chiral $\mathcal{S}$ and particle-hole (or charge conjugation) $\mathcal{C}=\mathcal{T}\mathcal{S}$ symmetries~\cite{winklerzulickeanziam2015}.
Time reversal is the antiunitary operation
\begin{equation}
\mathcal{T}: \Psi \rightarrow \alpha_yT^2 \mathcal{K} \Psi, \,\,\v{k}\rightarrow-\v{k},
\end{equation}
where $\mathcal{K}$ denotes complex conjugation.
Chiral symmetry $\mathcal{S}$ is a property of any nearest neighbor hopping fermionic model on a bipartite lattice.
It is the operation under which $c_A\rightarrow c_B^\dagger$ and $c_B \rightarrow -c_A^\dagger$, which here is encoded by
\begin{equation}
\mathcal{S} : \Psi \rightarrow \alpha_z T^3 \Psi.
\end{equation}
Finally, the particle-hole operation is
\begin{equation}
  \mathcal{C}: \Psi \rightarrow \alpha_x T^1 \mathcal{K}\Psi,\,\,\v{k}\rightarrow -\v{k}.
  \label{particlehole}
\end{equation}
It is important to note that additional real hopping terms, such as the next-nearest neighbor $t_2$ would break the particle-hole symmetry.
However at half-filling, particle-hole symmetry remains an emergent symmetry of the non-interacting low-energy Dirac theory.
This is because the additional term $\sim t_2 \abs{\v{k}}^2\alpha_0$ in the Hamiltonian, which breaks particle-hole symmetry, is of quadratic 
order in $\v{k}$ and hence an irrelevant perturbation in naive scaling analysis.

\subsection{Lattice Symmetry Breaking Charge Order} \label{seclatticesymmetrybreaking}
In recent years it has become well established~\cite{capponiiop2017,christouetalprb2018} that sufficiently strong next-nearest neighbor $V_2$ repulsions stabilize CDW$_3$ order on the honeycomb lattice, as is shown in Fig.~\ref{phasediagram}.
This charge ordering has reduced translational symmetry with a threefold enlarged unit cell, containing an entire hexagonal plaquette. 
Equivalently, the charge order has finite wavevector that couples the Dirac points, resulting in a threefold reduced Brillouin Zone.  Here we derive the effective GNY-like theory for charge ordering with reduced translational symmetry.
This is achieved by means of the Hubbard-Stratonovich transformation in the charge channel that introduces the six auxiliary dynamical charge order parameter fields
\begin{equation}
  \v{\rho} =(\rho_{A_1},\rho_{A_2},\rho_{A_3},\rho_{B_1},\rho_{B_2},\rho_{B_3} ),
\end{equation}
coupled to the density $\hat{n}_i$ on each site of the hexagon.
In light of the the extensive literature~\cite{capponiiop2017,christouetalprb2018}, we assume that the global minimum of the free energy has negligible weight in the bond order channel conjugate to $c^\dagger_i c_{j\ne i}$.

\subsubsection{Hubbard-Stratonovich Decomposition}
\begin{figure}[t]
 \includegraphics[width=\columnwidth]{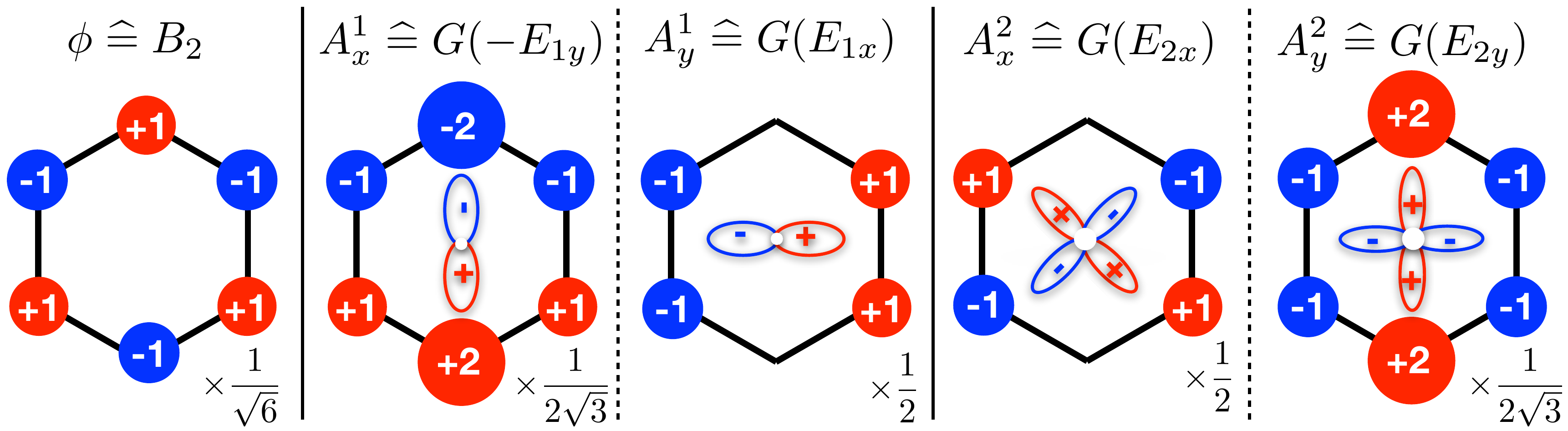}
 \caption{All possible charge instabilities in the 6-site unit cell
   on the half-filled honeycomb lattice.  Charge modulation is shown relative
   to half-filling. They are labelled with the order parameter fields that
 couple to the Dirac fermions in the effective field theory, as well
 as the corresponding elements in symmetry groups
 $C_{6v}^{\prime\prime}$ and $C_{6v}$ in parentheses.}
\label{plaquette}
\end{figure}

Here we present an overview of the steps taken to derive the low-energy effective field theory.
Further technical details can be found in Appendix \ref{appendixderivation}.
First, the interaction is decoupled in the charge channel,
\begin{equation}
\label{HS}
\begin{aligned}
&\exp\left[- \int_\tau \left( V_1\sum_{\avg{i,j}} \hat{n}_i \hat{n}_j+V_2 \sum_{\avg{\avg{i,j}}} \hat{n}_i \hat{n}_j \right)\right]\\
&= \int D[\v{\rho}] \exp \left[ - \int_\tau \sum_{\v{k}\in \text{RBZ}} \left(
  \v{\rho}_{\v{k}}^\dagger \, \v{V}_{\v{k}} \,\v{\rho}_{\v{k}} + 2
\hat{\v{n}}_{\v{k}}^\dagger\,\v{V}_{\v{k}}\,\v{\rho}_{\v{k}} \right)\right], \notag
\end{aligned}
\end{equation}
where the momentum summation is over the reduced Brillouin zone (RBZ) that is reciprocal to the enlarged unit cell.
The complicated effective interaction matrix $\v{V}_{\v{k}}$, that contains $V_1$ and $V_2$ terms, is defined in Eq.~\eqref{vmatrix}.
Then, the high energy modes up to $v_F \Lambda$ are integrated out by projecting into the Dirac subspace \eqref{diracsubspace}, $\Psi_i = \sum_{j} P_{ij} c_j$, 
where the $j$ sum is over the sites of the hexagon, and $P$ is defined in Eq.~\eqref{projection}.
Next, it is convenient to introduce the order parameter fields 
\begin{equation}
(\rho_0,\phi,A_x^1,A_y^1,A_x^2,A_y^2)= \v{U}\cdot \v{\rho},
\end{equation}
where
\begin{equation}
\v{U}=\frac{1}{6} \begin{pmatrix}
 \sqrt{6} & \sqrt{6} & \sqrt{6} & \sqrt{6} & \sqrt{6} & \sqrt{6} \\
 \sqrt{6} & \sqrt{6} & \sqrt{6} & -\sqrt{6} & -\sqrt{6} & -\sqrt{6} \\
 \sqrt{3} & -2 \sqrt{3} & \sqrt{3} & -\sqrt{3} & 2 \sqrt{3} & -\sqrt{3} \\
 -3 & 0 & 3 & 3 & 0 & -3 \\
 -3 & 0 & 3 & -3 & 0 & 3 \\
 -\sqrt{3} & 2 \sqrt{3} & -\sqrt{3} & -\sqrt{3} & 2 \sqrt{3} & -\sqrt{3}
\end{pmatrix}.
\end{equation}
The half-filling condition constrains the total charge density over the hexagonal plaquette, hence $\rho_0$ is fixed and thus neglected in the following. 
The corresponding charge patterns on the hexagonal unit cell that are induced by 
\begin{equation} 
\v{A} = (\v{A}^1,\v{A}^2),\,\,\v{A}^1=(A_x^1,
A_y^1),\,\,\v{A}^2=(A_x^2, A_y^2),
\end{equation}
%
and $\phi$  are shown in Fig.~\ref{plaquette}, from which the elements of $\v{U}$ can be read off.

In fact, the terms we have obtained are precisely the irreducible representations of the point group 
\begin{equation}
C^{\prime\prime}_{6v} = C_{6v}  + t_{\v{a}_1} C_{6v} +  t_{\v{a}_2} C_{6v},
\end{equation}
that contains the primitive translations $t_{\v{a}_{1,2}}$ \eqref{translations}, which has been discussed extensively in Refs.~\cite{baskoprb2008,dejuanprb2013}.
Here $\phi$ transforms as the $B_2$ irreducible representation and $(\v{A}^1,\v{A}^2)$ transform like the four components of $G$.
The components mix into each other under primitive translations  $\Psi \rightarrow e^{\pm i 2\pi T^3/3} \Psi$, thereby breaking the translational symmetry of $C_{6v}$ (the primitive unit cell).
The fields $(A_x^1,A_y^1,A_x^2,A_y^2)$ are distinguished by their properties under reflection
\begin{equation}
  \label{reflections}
  \begin{aligned}
R_x&: \text{(even, odd, odd, even)},\\
R_y&: \text{(odd, even, odd, even)},
\end{aligned}
\end{equation} 
%
which can be seen by inspection of Fig.~\ref{plaquette}.

Finally, the effective local Yukawa Lagrangian that couples Dirac fermions and (dynamical) order parameter fields has the compact form
\begin{equation}
L_g = \Psi^\dagger \left(g_\phi \phi \alpha_zT^3 + g_A \v{\alpha}\cdot\v{A}^aT^a\right)\Psi, \label{yukawa}
\end{equation}
where the implicit summation over repeated $a=1,2$ is used throughout. $L_g$ preserves emergent spatial rotational and $U(1)$ chiral symmetries, but breaks breaks Lorentz invariance.

The bare effective couplings are related to the lattice couplings,  $(g_\phi)_0 =\sqrt{3/2}(V_1-2V_2)$ and $(g_A)_0 = \sqrt{3}/2V_2$, but these will flow under the RG.
In addition, we obtain the bare Hubbard-Stratonovich order parameter mass term 
\begin{equation}
L_{\text{HS}} = \sqrt{3/2} (g_\phi)_0
\phi^2 + \sqrt{3} (g_A)_0 A^2, \label{hsmass}
\end{equation}
where $A=\sqrt{\v{A}^a\cdot\v{A}^a}$.    
The unimportant bare gradient terms have been suppressed in the interest of brevity.

\subsubsection{Mass Sector}
The anticommutator relation $\{ \hat{H}_0,  g_\phi \phi  \alpha_zT^3 \}=0$ identifies $\phi$ as a mass field that fully gaps the quasiparticle spectrum upon condensation~\cite{herbutetalprb2009b,ryuetalprb2009},
\begin{equation}
\varepsilon = \pm \sqrt{ \abs{v_F\v{k}}^2 + \avg{g_\phi\phi}^2},
\end{equation}
and is therefore expected to maximize the condensation energy gain.
In the regime $V_1\gg V_2$ the order parameter $\phi$ describes the quantum phase transition from the semimetal into the CDW insulator, which spontaneously breaks the sublattice exchange symmetry  $\Psi \rightarrow R_x \Psi$, $\phi \rightarrow - \phi$.
This corresponds to $Z_2$ chiral symmetry breaking in the low-energy effective field theory~\cite{gehringprd2015}, where the dynamics of $\phi$ are encapsulated by the Lagrangian
\begin{equation}
L_\phi = \frac{1}{2} \phi\left(-\partial_\tau^2-c_\phi^2\ \v{\partial}^2+m_\phi^2\right)\phi.
\end{equation} 

The order parameter mass $m_\phi^2$ tunes through the fermionic quantum critical point at $m_\phi^2=0$, which belongs to the $Z_2$-Gross-Neveu (or chiral  Ising) universality class.
It is understood to describe the same universal behavior as tuning through $V_1=(V_1)_c$, $(V_1\gg V_2)$ in the lattice model Eq.~\eqref{lattice}.
For $m_\phi^2<0$ the order parameter expectation value $\avg{\phi}$ is finite corresponding to $V_1>(V_1)_c$, where as for $m_\phi^2<0$ the symmetry is unbroken $\avg{\phi}=0$ corresponding to $V_1<(V_1)_c$.
The properties of this critical fixed point have been intensively studied. Further details can be found in Refs.~\cite{mihailaetalprb2017,zerfetalprd2017} and references therein.

For clarity, we now briefly discuss the remaining possible mass terms of four-component spinless Dirac fermions and their meaning on the honeycomb lattice.
The mass terms can be enumerated by
\begin{equation}
L_\text{mass} = \sum_{i=0}^{3} \Psi^\dagger M^i \alpha_z T^i \Psi.
\label{lmass}
\end{equation}
We reiterate that only the sublattice CDW mass~\cite{semenoffprl1984} $(M^3=\phi)$ is relevant in our paper.
The other terms arise from bond order $\avg{c^\dagger_i c_{j\ne i}}$ on the honeycomb lattice, and are not energetically favorable in the region of the phase diagram of interest $V_2\gg V_1$. 
Although at mean-field the Haldane~\cite{haldaneprl1988} quantum anomalous Hall mass $\Psi^\dagger M^0\alpha_z T^0 \Psi $ is stable  for $V_2\gg V_1$~\cite{raghuetalprl2008}, upon the inclusion of beyond mean-field quantum fluctuations it is known that CDW$_3$ order is favored~\cite{capponiiop2017,christouetalprb2018}.
The masses $\Psi^\dagger(M^1\alpha_zT^1,M^2\alpha_z T^2)\Psi$ form an XY order parameter that corresponds to the Kekul\'{e} valence bond solid phase~\cite{houetalprl2007,lietalnatcom2017,schererherbutprb2016}, which is understood to be the leading instability for strong $V_1\approx V_2$.

\subsubsection{Emergent Gauge Sector}
Following the discussion of the mass fields, it is important to notice that all possible remaining terms that minimally couple to four-component Dirac fermions are gauge fields,
\begin{equation}
L_\text{gauge} = \sum_{i=0}^{3} \sum_{j=0,x,y} \Psi^\dagger A^i_j \alpha_j T^i\Psi,
\label{lgauge}
\end{equation}
where $A_j^0$ are the Abelian and $A_j^{i>0}$ the non-Abelian components. From this identification it follows that any order parameter associated with 
lattice symmetry breaking is necessarily composed of a combination of mass
and emergent gauge fields.  Although we have focused on the honeycomb lattice, it is evident that this is a generic property of Dirac materials.
We now explicitly demonstrate that this is the case for CDW$_3$ order. 

In the low-energy, $\v{A}^{\alpha}$ $(\alpha=1,2)$ minimally couple as components of an emergent $SU(2)$
non-Abelian local gauge theory generated by $T^a$.
This is revealed by the local Lagrangian
\begin{equation}
L_\Psi + L_{g_A} = \Psi^\dagger \left[\partial_\tau + i\v{\alpha}\cdot
 \left(v_F \v{\partial} - i g_A \v{A}^a T^a\right) \right] \Psi,
\end{equation}
which is invariant under the local gauge transformation generated by $T^{1,2}$.
To avoid integrating over the emergent gauge redundancies, the theory is gauge fixed in the $R_\xi$ gauge by following the Fadeev-Popov procedure~\cite{peskin1996introduction}.
The dynamics of $\v{A}^{1,2}$ are described by the gauge fixed Lagrangian
\begin{align}
L_A&=\frac{1}{2}\v{A}^a \cdot\left(-\partial_\tau^2-c_A^2\v{\partial}^2+m_A^2\right) \v{A}^a\notag\\
&\phantom{=}\,+\frac{\xi-1}{2\xi}\sum_{i,j=x,y}A_i^ac_A^2\partial_i\partial_j A^a_j ,
\end{align}
which is invertible for all finite $\xi$.
It is convenient to employ the Feynman-'t Hooft gauge $\xi=1$, where the Lagrangian is spatially isotropic.
Naturally, the universal behavior should not depend on the choice of gauge.

The bosonic mass $m_A^2$ is finite from the lattice Hubbard-Stratonovich transformation \eqref{hsmass} with the bare mass $(m_A^2)_0= 3 V_2$.
This implies that $\v{A}^{1,2}$ can have finite order parameter expectation values~\cite{kimetalprb2008,huhsachdevprb2008,metlitskisachdevprb2010}, which is of course expected from the lattice model where $\v{A}^{1,2}$ are linear combinations of charge order parameters.
This is in contrast to the standard Yang-Mills theory, where $m_A^2=0$ is enforced by the Ward identity associated with the chiral/gauge invariance \eqref{chiralgaugesymmetry}.

\subsubsection{Order Parameter Self-Interactions} 
From symmetry considerations, or alternatively by integrating out high energy fermionic modes, one obtains the self-interaction of the dynamical order parameter fields
\begin{equation}
  L_\lambda=\lambda_{\phi}  \phi^4+ \lambda_{A} A^4
+\lambda_{\phi A}\phi^2 A^2 +\lambda_{\text{YM}}  (\v{A}^1\times\v{A}^2)^2.
\label{quartic}
\end{equation}
The Yang-Mills term $\lambda_{\text{YM}}$ reflects the underlying emergent non-Abelian gauge structure.
The coupling $\lambda_{\phi  A}$ implies that the condensation of the mass field $\phi$ could dynamically generate a finite expectation value for the emergent gauge fields, 
which is reminiscent of a Higgs mechanism.

Additionally, in the low-energy theory there is the symmetry allowed analytic cubic term
\begin{equation}
L_{\tilde{b}_3}=  \tilde{b}_3\phi  (\v{A}^1\times\v{A}^2)_z.
\label{b3t}
\end{equation}
This is because $\v{A}^1\times\v{A}^2$ preserves the emergent global chiral symmetry generated by $T^3$, under which
\begin{equation}\label{Gtranslation}
  \begin{pmatrix}
\v{A}^1\\\v{A}^2
\end{pmatrix}
\rightarrow
\begin{pmatrix}
  \cos\theta&&-\sin\theta\\
  \sin\theta&&\cos\theta
\end{pmatrix}
  \begin{pmatrix}
\v{A}^1\\\v{A}^2
\end{pmatrix}.
\end{equation}
Likewise the spatial rotational symmetry \eqref{rotationalsymmetry} generated by $\alpha_z$ is preserved, under which
\begin{equation}\label{Grotation}
  \begin{pmatrix}
A_x^a\\A_y^a
\end{pmatrix}
\rightarrow
\begin{pmatrix}
  \cos\theta&&-\sin\theta\\
  \sin\theta&&\cos\theta
\end{pmatrix}
  \begin{pmatrix}
A_x^a\\A_y^a
\end{pmatrix}.
\end{equation}
Yet, the non-interacting theory is symmetric under the particle-hole transformation \eqref{particlehole}, where as the Yukawa terms in $L_g$ \eqref{yukawa}  are all particle-hole odd.
Therefore the renormalization of the cubic terms by fermion loops are forbidden, as all loop corrections vanish by symmetry,
\begin{align}
&g_\phi g_A^2 \phi A_i^a A_j^b\avg{\Psi^\dagger \alpha_z T^3 \Psi \Psi^\dagger \alpha_i T^a \Psi \Psi^\dagger \alpha_j T^b \Psi }\rightarrow\notag\\ &-g_\phi g_A^2 \phi A_i^a A_j^b\avg{\Psi^\dagger \alpha_z T^3 \Psi \Psi^\dagger \alpha_i T^a \Psi \Psi^\dagger \alpha_j T^b \Psi }=0.
\end{align}
This is in accordance with Furry's theorem~\cite{peskin1996introduction}, and extends to all fermion loops with an odd number of external legs. However, $\tilde{b}_3$ may still be renormalized by self-interactions.

\subsubsection{Lattice Symmetry Allowed Self-Interactions}
The reduced symmetry of the honeycomb lattice $C_{6v}^{\prime\prime}$ allows for additional interactions.
Such terms are not generated (or renormalized) by the Dirac fermion loops, as they possess the higher continuous spatial rotational symmetry.

These terms are best exposed by introducing the double-complex, Hopf-coordinate representation

\begin{equation}
  G= G_1 + j G_2,\;\text{ with }\;G_a = A_x^a+iA_y^a.
\end{equation}

Here spatial rotations are implemented by $e^{i\theta}$ and primitive translations (chiral transformations) are implemented by $e^{j\theta}$.

The lattice symmetry allowed cubic self-interaction is identified by first decomposing $G^3$ into the real (R) and imaginary (I) parts in complex $i$ and $j$

\begin{equation}
G^3= (G^3)_{\text{RR}}+j(G^3)_{\text{IR}} + i(G^3)_{\text{RI}}+ ij(G^3)_{\text{II}}.
\end{equation}

Each of the four terms is individually invariant under primitive translations $G\rightarrow e^{j2\pi/3}G$ and $C_3$ rotations $G\rightarrow e^{i2\pi/3}G$. 
Imposing invariance under reflections in $x$ and $y$ planes \eqref{reflections} reduces the symmetry allowed term to 

\begin{align}
  L_{b_3}&=b_3(G^3)_{\text{II}}=b_3\text{Im}[G_2^3-3 G_1^2 G_2]\notag\\
                  &= b_3\left\{3 \left[(A_y^1)^2-  (A_x^1)^2 +( A_x^2 )^2\right] (A_y^2) \right. \notag\\
         &\left.\phantom{=}\, -6 (A_x^1) (A_x^2) (A_y^1) -( A_y^2)^3\right\}.
\label{b3}
\end{align}

Repeating this type of analysis for $\phi G^3$ identifies the lattice symmetry allowed quartic self-interaction

\begin{equation}
L_{b_4}=b_4\phi\,\text{Re}[G_1^3-3 G_1 G_2^2].
\end{equation}

The cubic terms $L_{\tilde{b}_3}$ \eqref{b3t} and $L_{b_3}$ \eqref{b3} could potentially render the transition first order. While such terms would be 
relevant at the Wilson-Fisher fixed point, we will show in Sec.~\ref{sectionirrelevantcubic} that they vanish at the fermion-induced critical fixed point ~\cite{lietalnatcom2017}.

\section{Renormalization Group Analysis}
\label{secrgequations}
Quantum (multi-)critical points are described by scale invariant fixed points of the RG transformation 
\begin{equation}
 k_0 = k^\prime_0 e^{-z \ell},\,\,k_x = k^\prime_x e^{-\ell},\,\,k_y = k^\prime_y e^{-\ell}.
\end{equation}
We use the perturbative Wilson momentum shell scheme, in which we integrate over fast momentum modes 
\begin{equation}
\Lambda  e^{-\ell}\le \abs{\v{k}} \le \Lambda,\,\, -\infty\le k_0\le\infty,
\end{equation}
to identify universal features of the action $S=\int_{\tau,\v{r}}L$ for
\begin{equation}
  L=L_\Psi + L_\phi+L_A +L_g+L_\lambda.
\end{equation}
Following this main analysis, we will then analyze the role of the symmetry allowed self-interactions 
\begin{equation}
L_b=L_{b_3}+L_{b_4}+L_{\tilde{b}_3}, \label{Lbequation}
\end{equation}
which play a secondary role due to particle-hole and rotational symmetries preventing renormalization from the gapless fermions.

Criticality is accessed by tuning the order parameter masses to their fixed point values $(m_{\phi,A}^2)_*$. 
We follow the usual procedure and absorb the small linear shifts from vertex corrections into a re-definition of the masses such that 
$(m_{\phi,A}^2)_*=0$. The masses are relevant perturbations at any fixed point and can be tuned by the lattice interaction parameters.

\subsection{Dimensional Continuation of the Lagrangian}
Tree-level scaling of GNY theories determines that the dynamical exponent is equal to $z=1$, and that the Yukawa and self-interactions are marginal perturbations at the Gaussian fixed point in $d=3$ spatial dimensions.
This motivates an $\epsilon=3-d$ expansion to access the strongly interacting quantum critical point in a controlled manner.
Further control is exerted by generalizing to a large number $N$ of fermionic spin flavors,
\begin{equation}
	\Psi^\dagger\Psi \rightarrow
       \sum_{n=1}^N\Psi^\dagger_n\Psi_n,\;\;
      (g^2_{\phi,A}, \lambda_i) \rightarrow \frac{8\pi^2 \Lambda^\epsilon}{N}(g^2_{\phi,A},\lambda_i),
    \end{equation}
which is known to favorably reorganize the perturbative expansion, enabling perturbative RG directly in the physically relevant spatial dimension $d=2$~\cite{moshezinnjustinpr2003}. 
Here $i$ indexes the couplings in Eq.~\eqref{quartic}.

The dimensional continuation to $d=3-\epsilon$ spatial dimensions (for
$\epsilon>0$ or $d<3$) is best formulated with the anticommuting Dirac
$\gamma$ matrices defined below Eq.\eqref{psigamma} and
\begin{equation}
(\gamma_3,\gamma_5,\gamma_{35})=(T^1,T^2,T^3),
\end{equation}
where $\gamma_{35}=-i \gamma_3\gamma_5$ and with $\{\gamma_i,\gamma_j\} =2\delta_{ij}$ for $i,j =
0,\dots, 5$.
Then the Lagrangian in continuous spatial dimension $1<d<3$ is
\begin{align}
L&=\sum_{n=1}^{N}\bar{\Psi}_n \bigg[\partial_\tau \gamma_0 + v_F \partial_\mu \gamma_\mu
   + \frac{g_\phi }{\sqrt{\tilde{N}}} \phi \notag\\
  &\phantom{=}\,- \frac{i g_A }{\sqrt{\tilde{N}}} \left( \gamma_1 \gamma_3 A_x^1 + \gamma_2 \gamma_3 A_y^1 + \gamma_1 \gamma_5 A_x^2 + \gamma_2 \gamma_5 A_y^2\right)\bigg]\Psi_n\notag\\
 &\phantom{=}\,+\frac{1}{2} \phi\left(-\partial_\tau^2-c_\phi^2\partial^2_\mu+m_\phi^2\right)\phi
\notag \\
  &\phantom{=}\, +\frac{1}{2}\v{A}^a
              \cdot\left(-\partial_\tau^2-c_A^2\partial^2_\mu+m_A^2\right)
              \v{A}^a
 \notag \\ 
 &\phantom{=}\, + \frac{1 }{\tilde{N}}\left[\lambda_{\phi}  \phi^4+ \lambda_{A} A^4
              +\lambda_{\phi A}\phi^2 A^2 +\lambda_{\text{YM}}  (\v{A}^1\times\v{A}^2)^2\right],
\end{align}
where $\tilde{N}= N/8\pi^2 \Lambda^\epsilon$ and there is the implicit summation over repeated $\mu=1,\dots,d$. In this case $\text{Tr}(\gamma_\mu\gamma_\mu)=4d$.
The corresponding fermion propagator is
\begin{equation}
G_\Psi(k_0,k_\mu) = \frac{i (k_0 \gamma_0+ v_F k_\mu \gamma_\mu)}{k_0^2 + v_F^2 k_\mu^2},
\label{fermionpropagator}
\end{equation}
and the order parameter propagators are 
\begin{equation}
G_{\phi,A}(k_0,\v{k}) = \frac{1}{k_0^2 + c_{\phi,A}^2 k_\mu^2+m_{\phi,A}^2}.
\label{bosonpropagator}
\end{equation}
Under the space-time rescaling the fermion and order parameter fields rescale as
\begin{align}
\Psi(k_0,k_\mu)&=\Psi^\prime(k_0^\prime,k_\mu^\prime)e^{(2z+d-\eta_\Psi)\frac{\ell}{2}},\\
\phi(k_0,k_\mu)&=\phi^\prime(k_0^\prime,k_\mu^\prime) e^{(3z+d-\eta_\phi)\frac{\ell}{2}},\\
\v{A}^a(k_0,k_\mu)&={\v{A}^a}^\prime(k_0^\prime,k_\mu^\prime) e^{(3z+d-\eta_A)\frac{\ell}{2}},
\end{align}
where $\eta_{\Psi,\phi,A}$ are the anomalous dimensions that account for the beyond tree-level corrections to the field rescaling. Our convention is that the rescaling of the fields ensures the scale invariance of the imaginary time gradients.

\begin{figure}[t]
 \includegraphics[width=0.4\textwidth]{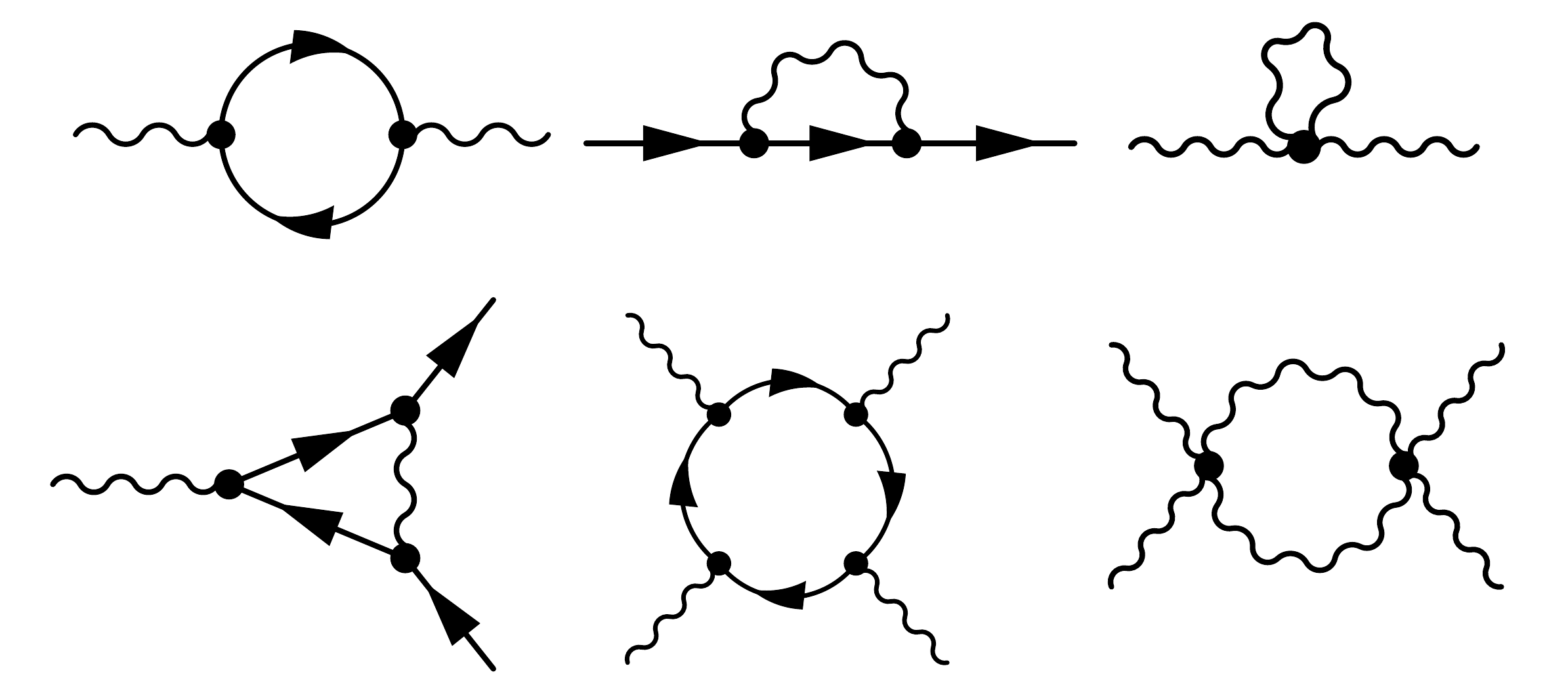}
 \caption{One-loop Feynman diagrams. The fermion propagator is denoted by the arrowed line. The boson propagators are denoted by the wavy line.}
\label{diagrams}
\end{figure}

\subsection{Renormalization Group Equations}
The RG equations are obtained to one-loop order by calculating the
diagrams in Fig.~\ref{diagrams}, using the critical $(m_{\phi,A}^2=0)$ order parameter propagators \eqref{bosonpropagator}.
Scale invariance implies the anomalous dimensions are
\begin{equation}
\begin{gathered}
   \eta_\phi =\frac{2g_\phi^2}{ v_F^3},\,\,
      \eta_A =\frac{4g_A^2}{3 v_F^3},\\
\eta_\Psi  =\frac{2}{N} \bigg[\frac{ g_{\phi }^2 }{
   c_{\phi } \left(c_{\phi }+v_F\right){}^2}   + \frac{
   4g_A^2 }{c_A \left(c_A+v_F\right){}^2}\bigg].
      \end{gathered}
\end{equation}
Then the RG equations for the velocities are
  \begin{align}
\dl v_F &= (z-1) v_F+ \frac{4}{N}\frac{g_{\phi }^2 \left(c_{\phi }-v_F\right)}{3
   c_{\phi } \left(c_{\phi }+v_F\right){}^2}
   \notag \\ &\phantom{=}
   -\frac{4}{N}\frac{2
   g_A^2 v_F}{c_A \left(c_A+v_F\right){}^2}\label{fermionvelocity},\\
 \dl c_\phi^2 &=2 (z-1) c_{\phi }^2+\frac{2g_{\phi }^2 \big(v_F^2-c_{\phi }^2\big)}{
   v_F^3}\label{bosonvelocity1}, \\
  \dl  c_A^2&=2 (z-1)
   c_A^2+\frac{4g_A^2 \big(v_F^2-c_A^2\big)}{3 v_F^3}\label{bosonvelocity2}.
  \end{align}
   It is important to highlight that divergent anisotropic Fermi velocity renormalization, preempting a fixed point, is encountered if the independent spatial 
   rotational symmetries of $\v{A}^1$ or $\v{A}^2$ are artificially broken.
   The RG equations for the Yukawa couplings are
   \begin{align}
  \dl g_\phi^2 &= g_{\phi }^2 \bigg[\epsilon + 3 (z-1) -\frac{2g_{\phi
   }^2}{ v_F^3}-\frac{4}{N}\frac{g_{\phi }^2
   \left(c_{\phi }+2 v_F\right)}{c_{\phi } v_F
   \left(c_{\phi }+v_F\right){}^2} \notag \\ & \phantom{=g_{\phi }^2 \bigg[}\,
   +\frac{4}{N}\frac{4 g_A^2 \left(c_A+2
   v_F\right)}{c_A v_F
   \left(c_A+v_F\right){}^2}\bigg],\label{yukawa1}\\
\dl g_A^2  &= g_A^2 \bigg[\epsilon+3(z-1)-\frac{4g_A^2}{3
   v_F^3}-\frac{4}{N}\frac{2 g_{\phi }^2
   \left(c_{\phi }+2 v_F\right)}{3 c_{\phi } v_F
   \left(c_{\phi }+v_F\right){}^2}
   \notag \\ & \phantom{=g_{\phi }^2 \bigg[}\,
+   \frac{4}{N}\frac{4 g_A^2}{c_A
   \left(c_A+v_F\right){}^2}\bigg].\label{yukawa2}
   \end{align}
   The RG equations for the order parameter self-interactions are
   \begin{align}
 \dl \lambda_\phi  &=\lambda _{\phi} \bigg[\epsilon+
   3(z-1)-\frac{4g_{\phi }^2}{v_F^3}\bigg]
  +\frac{g_{\phi }^4}{ v_F^3} 
\notag \\  & \phantom{=}\, 
  -\frac{4}{N}\bigg[\frac{9
   \lambda _{\phi }^2}{c_{\phi }^3}+\frac{\lambda _{\phi  A}^2}{c_A^3}\bigg], \label{self1}\\
 \dl \lambda_A &= \lambda _A
\bigg[\epsilon+3(z-1)-\frac{8 g_A^2}{3 v_F^3}\bigg]
 \notag \\  & \phantom{=}\,
   -\frac{4}{N}\bigg[\frac{12 \lambda _A^2}{c_A^3}+\frac{\lambda
   _{\text{YM}}^2}{4 c_A^3}+\frac{\lambda _{\phi  A}^2}{4 c_{\phi
   }^3}
   +\frac{\lambda _A
   \lambda _{\text{YM}}}{c_A^3}\bigg], \\
   \dl \lambda_{\phi A}&=\lambda
   _{\phi  A} \bigg[\epsilon+3(z-1)-\frac{4g_A^2}{3
   v_F^3}-\frac{2g_{\phi }^2}{ v_F^3}\bigg]+\frac{4g_A^2 g_{\phi }^2}{v_F^3} \notag \\
   &\phantom{=}\,-\frac{4}{N}\bigg[\frac{4 \lambda _{\phi  A}^2}{c_A
   c_{\phi } \left(c_A+c_{\phi }\right)}+\frac{3 \lambda _{\phi }\lambda _{\phi  A} }{c_{\phi
   }^3}
   \notag \\  & \phantom{=-\frac{4}{N}\bigg[}\, 
   +\frac{6 \lambda
   _A\lambda _{\phi  A} }{c_A^3}+\frac{\lambda _{\text{YM}}\lambda _{\phi  A} }{2
   c_A^3}\bigg],\\ 
  \dl \lambda_{\text{YM}}&= \lambda
   _{\text{YM}} \bigg[\epsilon + 3(z-1)-\frac{8 g_A^2}{3 v_F^3}\bigg]+\frac{8 g_A^4}{3 v_F^3}   \notag \\  & \phantom{=}\,-\frac{4}{N}\bigg[\frac{2 \lambda _{\text{YM}}^2}{c_A^3}+\frac{12
   \lambda _A \lambda _{\text{YM}}}{c_A^3}\bigg].\label{self4}
  \end{align}
    The RG equations for the boson order parameter field masses are
  \begin{align}
   \dl m^2_\phi&=m_{\phi }^2 \bigg[2
   z-\frac{2g_{\phi }^2}{ v_F^3}-\frac{12 \lambda _{\phi } }{N c_{\phi }^3}\bigg]-m_A^2\frac{8 \lambda _{\phi 
   A}}{N c_A^3},               \label{rgmass1}\\
  \dl m^2_A &=m_A^2 \bigg[2
   z-\frac{4g_A^2}{3 v_F^3}-\frac{2 \left(12 \lambda _A+\lambda
   _{\text{YM}}\right)}{N c_A^3}\bigg]  \notag \\  & \phantom{=}\,
-m_{\phi }^2 \frac{2\lambda _{\phi  A}}{N c_{\phi
              }^3}.
              \label{rgmass2}
  \end{align}
  Note that to obtain Eqs.~\eqref{rgmass1} and \eqref{rgmass2} the propagators must contain finite order parameter masses, and the diagrams are then expanded to leading order.
   
The RG equations are solved to $\mathcal{O}(1/N)$ and to leading order in $\epsilon$ as follows. First, $z$ is chosen to make $v_F$ scale invariant by solving Eq.~\eqref{fermionvelocity}. Then, the fixed points of the boson velocities and Yukawa couplings are obtained simultaneously by solving Eqs.~(\ref{bosonvelocity1}\,--\,\ref{yukawa2}). 
The boson self-interaction fixed points are then obtained by solving Eqs.~(\ref{self1}\,--\,\ref{self4}).

\begin{table*}[t] 
\center
\begin{tabular}{c | c | c | c | c}
 Exponent &GNY  $(d=3-\epsilon)$ & CDW$_3$  $(d=3-\epsilon)$ & CDW$_3$ $(d=2)$ & CDW$_A$ $(d=3-\epsilon)$ \\ \hline
$ z-1 $ & $0$ & $\frac{3\epsilon}{2N}$ &$\frac{56 \sqrt{2}-75}{N}$&$\frac{3\epsilon}{2N} $\\
 $ \eta_\Psi$ &$\frac{\epsilon}{4N}$&$\frac{7 \epsilon}{4N}$&$\frac{54 \sqrt{2} -72}{N}$&$\frac{3 \epsilon}{2N}$\\
 $\eta_{\phi}$ &$\epsilon\left(1-\frac{3}{2N}\right)$&$\epsilon\left(1-\frac{6}{N}\right)$&$1+\frac{24\sqrt{2} -45}{N}$&\\
 $\eta_{A}$ & $$& $\epsilon\left(1+\frac{1}{2N}\right)$& $1+\frac{64\sqrt{2}-87}{N}$& $\epsilon\left(1+\frac{3}{2N}\right)$\\
 $2-\nu^{-1}_{1}$&$ \epsilon\left(1+\frac{3}{2N}\right)$&$ \epsilon\left(1+\frac{11+\sqrt{745}}{4N}\right)$  & $1+ \frac{84 -65\sqrt{2} + \sqrt{1307-714\sqrt{2}} }{N}$&$\epsilon\left(\frac{3}{5} + \frac{9}{10 N}\right)$\\
  $2-\nu^{-1}_{2}$&$$&$ \epsilon\left(1+\frac{11-\sqrt{745}}{4N}\right)$  & $1+ \frac{84 -65\sqrt{2} - \sqrt{1307-714\sqrt{2}} }{N}$&
\end{tabular}
\caption{One-loop critical exponents at the GNY/Chiral Ising,  CDW$_3$, and CDW$_A$ fixed points to $\mathcal{O}(1/N^2)$. 
The GNY fixed point at $m_\phi^2=0, m_A^2\gg 0$ describes the quantum critical point of the semimetal-insulator transition into the sublattice CDW phase on the honeycomb lattice. 
The CDW$_3$ fixed point at $m_\phi^2=0,m_A^2=0$ describes a new fixed point of interacting Dirac fermions where lattice symmetries are spontaneously broken.
For the CDW$_3$ fixed point there is good agreement between the $\epsilon=3-d$ expansion and the direct evaluation in the physical  dimension $d=2$ as $N\rightarrow\infty$. 
The CDW$_A$ fixed point at $m_\phi^2>0, m_A^2= 0$ is shown for completion, but is not expected to be physically accessible from the 
microscopic model on the honeycomb lattice \eqref{lattice}.}
    \label{exptab}
  \end{table*}

The RG flow in the vicinity of the multicritical fixed point  $(m_{\phi,A}^2)_*=0$  is determined by linearization in the couplings $ x_i = (g^2_\phi,g^2_A,\lambda_\phi,\lambda_{A},\lambda_{\text{YM}},\lambda_{\phi A},m^2_\phi,m^2_A)$ around their fixed point values $(x_i)_*$. 
The inverse correlation length exponents $\nu^{-1}_{1,2}$ are determined by the  two relevant (positive) eigenvalues of the stability matrix 
\begin{equation}
X_{ij} = \frac{\partial}{\partial x_j} \dl x_i \bigg|_{x_i=(x_i)_*}.
\end{equation}

\subsection{Critical Fixed Points}
The Yukawa couplings are relevant perturbations at the putative Wilson-Fisher fixed points ($\{\lambda_i\} \ne \{0\}$, $g_{\phi,A}=0)$, owing to the gapless nature of the fermion excitations.
At the ensuing fermionic critical fixed points ($\{\lambda_i,g_{\phi,A}\} \ne \{0\}$) both the order parameter $\eta_{\phi,A}$ and fermion $\eta_\Psi$  anomalous dimensions are finite. 
The latter indicates the breakdown of the quasiparticle picture and the onset of a non-Fermi liquid.

However, the coupling to a particular fluctuating order parameter field can be rendered irrelevant if the related order parameter mass is tuned far from critical~\cite{zinnjustinnpb1991,janssenetalprb2018}.
In this case the order parameter field can be integrated out and the corresponding fermion-fermion interaction,  $V\sim \mathcal{O}(g^2/m^2)$, will be vanishingly small.
For example, in the case where $m_A^2\gg0$ and $m_\phi^2=0$, the theory describes the GNY/chiral Ising quantum critical fixed point of the CDW transition at $V_1 = (V_1)_c$ and $V_2 \ll (V_2)_c$. Similarly, there is the critical fixed point (CDW$_A$) at $m_A^2=0$ and $m_\phi^2\gg0$, although this is not expected to be physically accessible 
for the microscopic model on the honeycomb lattice \eqref{lattice}. The critical exponents are summarized in Tab.~\ref{exptab}.

\subsection{CDW$_3$ Criticality}
The CDW$_3$ fixed point is located at $m_\phi^2= 0$ and $m_A^2=0$. In the large $N$ limit there is a well defined multicritical fixed point
\begin{equation}
\begin{gathered}
 (g_\phi^2,g_A^2)_* = \epsilon(2-\frac{12}{N},3+\frac{3}{2N}),\\
(\lambda_\phi,\lambda_A,\lambda_{\phi A},\lambda_{\text{YM}})_* = \epsilon(1-\frac{81}{2N},-\frac{18}{N},6-\frac{81}{N},6-\frac{45}{N}).
\end{gathered}
 \label{criticalYukawa}
 \end{equation} 
The spontaneous lattice symmetry breaking (finite $g_A$) results in the violation of Lorentz invariance with 
 \begin{equation}
 z=1+\frac{(g_A^2)_*}{2N}= 1 +\frac{3\epsilon}{2N}.
 \end{equation}
This is in contrast to the emergent Lorentz invariance of GNY fixed points with  $z=1$ and $c_\phi=v_F$ to all orders in $N$.
The resulting critical exponents are collected in Tab.~\ref{exptab}, and are contrasted with the GNY universality.
The finite anomalous dimensions indicate the breakdown of the quasiparticle picture at the multicritical point.
There is good agreement from analysis of the CDW$_3$ fixed point directly in the physical spatial dimension $d=2$, which is perturbatively controlled by large $N$ (RG equations in Appendix \ref{appendixrgequations}).

\subsection{Cubic Terms and Fermion-Induced Criticality}\label{sectionirrelevantcubic}
Our RG analysis indicates the existence of a continuous quantum phase transition. 
However, there are symmetry allowed cubic terms in $L_b$ \eqref{Lbequation}. 
If the cubic couplings in $L_b$ at a putative fixed point are finite, the Landau cubic criterion could imply a first order transition. 

The naive tree-level scaling $[b_3]=[\tilde{b}_3]=(5z-d)/2$ suggests that this is indeed the case.
Yet, the large order parameter anomalous dimensions render the cubic term irrelevant at the $d=2$ CDW$_3$ critical point. 
Therefore the coupling to gapless Dirac fermions induces a continuous transition, an example of fermion-induced criticality~\cite{lietalnatcom2017}.

To demonstrate this it is sufficient to calculate the  RG equations to leading order,
\begin{align}
\dl b^2_3 &= b^2_3\left[ \epsilon + 5z-3  -
  \frac{4g^2_A}{v^3_F}-\frac{12}{N}\left(\frac{3\lambda_A}{c_A^3} +
    \frac{ \lambda_{\text{YM}}}{c_A^3}\right)\right],\\
   \dl \tilde{b}^2_3 &= \tilde{b}^2_3\bigg[ \epsilon + 5z-3  -
  \frac{8g^2_A+6g_\phi^2}{3v^3_F} -\frac{4}{ N}\bigg(
 \frac{ 4\lambda_A}{c_A^3}\notag \\
  &\phantom{=\tilde{b}^2_3\bigg[}\,+ \frac{3\lambda_{\text{YM}} }{c_A^3} + \frac{8 \lambda_{\phi A}}{c_\phi c_A (c_\phi + c_A)}\bigg)\bigg],
     \label{brge}
\end{align}
where the couplings are rescaled, $b^2_3\rightarrow 8\pi^2 \Lambda^\epsilon b^2_3/N$, and similar for $\tilde{b}_3^2$. 
This requires calculating the one-particle irreducible contractions of $\avg{\int L_b \int L_\lambda}_>$ with three slow order parameter legs.
 Crucially, there is no direct renormalization by the Dirac fermions because of their higher continuous spatial rotational symmetry and particle-hole symmetry, as discussed in Sec.~\ref{seclatticesymmetrybreaking}. 
 At the CDW$_3$ fixed point the RG equations reduce to 
\begin{align}
\dl b^2_3 &= 2 b^2_3\left[ 1-\epsilon-\frac{24\epsilon}{N}\right] ,\\
\dl \tilde{b}^2_3 &= 2\tilde{b}^2_3\left[1-\epsilon -  \frac{59\epsilon}{4N}\right].
\end{align}
Therefore, to this order the cubic terms are irrelevant perturbations in $d=2$ spatial dimensions $(\epsilon=1)$, and the fixed point value is $(b_3)_*=(\tilde{b}_3)_*=0$. 
In $N\rightarrow \infty$, the cubic terms are marginal to all loop orders following the arguments in Ref.~\cite{schererherbutprb2016}.
Similar behavior was found in the case of Kekul\'{e} ordering. 
For the complex Kekul\'{e} XY order parameter $\chi=(M^1+i M^2)/\sqrt{2}$ there is the lattice symmetry allowed Potts clock term $\chi^3 + {\chi^*}^3 \sim \abs{\chi}^3\cos(3\varphi)$, which is found to be irrelevant at the fermionic critical point~\cite{lietalnatcom2017}. 
This highlights that fermion-induced critical points are inherently different  from Wilson-Fisher fixed points of conventional order parameter theories. 
Naturally, quartic and higher order lattice  interactions are even more irrelevant at fermionic fixed points.

\section{Broken Symmetry State}\label{secbrokensymmetry}
\subsection{CDW$_3$ Broken Symmetry State}
To analyze the nature of the CDW$_3$ broken symmetry state, the free energy density $f(\phi,\v{A}^1,\v{A}^2)$ must be minimized.
Infinitesimally close to the multi-critical point, $f$ is obtained from integrating over the fermions for static order parameter fields with finite expectation values~\cite{moshezinnjustinpr2003}. Although the criticality is universal, the broken symmetry state is not. 
Instead it depends on the lattice model and the concomitant path taken through the critical surface $(\delta_\phi,\delta_A)=\delta(\cos\theta,\sin\theta)$ with $\delta_{\phi,A} = (m_{\phi,A}^2)_* - m_{\phi,A}^2$.

The analysis proceeds directly in $d=2$ and is  controlled with large $N$. 
In the region of the critical surface it is assumed that the couplings are well approximated by their fixed point values $(m_{\phi,A}^2)_* = 4 \Lambda^2$, $(g_\phi^2)_* = 4\pi \Lambda/N$, $(g_A^2)_*=2(g_\phi^2)_*$, and $v_F=1$. 
Integrating over the fermions, we obtain the Landau free energy density 
\begin{align}
f&= -N \int^{\infty}_{-\infty} \frac{d k_0}{2\pi} \int^\Lambda \frac{d^2\v{k}}{(2\pi)^2}  \ln \det \bigg(  -i k_0 + \v{\alpha}\cdot\v{k}  \notag \\
  &\phantom{=}\, + \frac{(g_\phi)_*}{\sqrt{N}} \phi \alpha_zT^3 +
\frac{(g_A)_*}{\sqrt{N}}  \v{\alpha}\cdot\v{A}^aT^a \bigg)\notag\\ 
&\phantom{=}\,+\frac{m_\phi^2}{2}\phi^2 +\frac{m_A^2}{2}A^2+\cdots.
\end{align}
Here $\cdots$ indicate higher order terms, which includes those that depend on the lattice model.
For $V_2\gg V_1$ on the honeycomb lattice, unconstrained charge order has an ill-defined Hubbard-Stratonovich transformation.
Therefore, it is expedient to constrain the order parameters space to the physically valid CDW$_3$ region, 
\begin{align}
\phi&=\sqrt{2/3} (\rho-\Delta),\\
(\v{A}^1,\v{A}^2)&=\frac{4\rho+2\Delta}{\sqrt{3}}(C_\alpha C_{\beta},C_\alpha S_{\beta},S_\alpha C_{\gamma},S_\alpha S_{\gamma}),
\end{align}
where $\rho\ge\Delta\ge0$, $C_\alpha=\cos\alpha$, $S_\alpha=\sin\alpha$.

\begin{figure}[t]
 \includegraphics[width=0.5\textwidth]{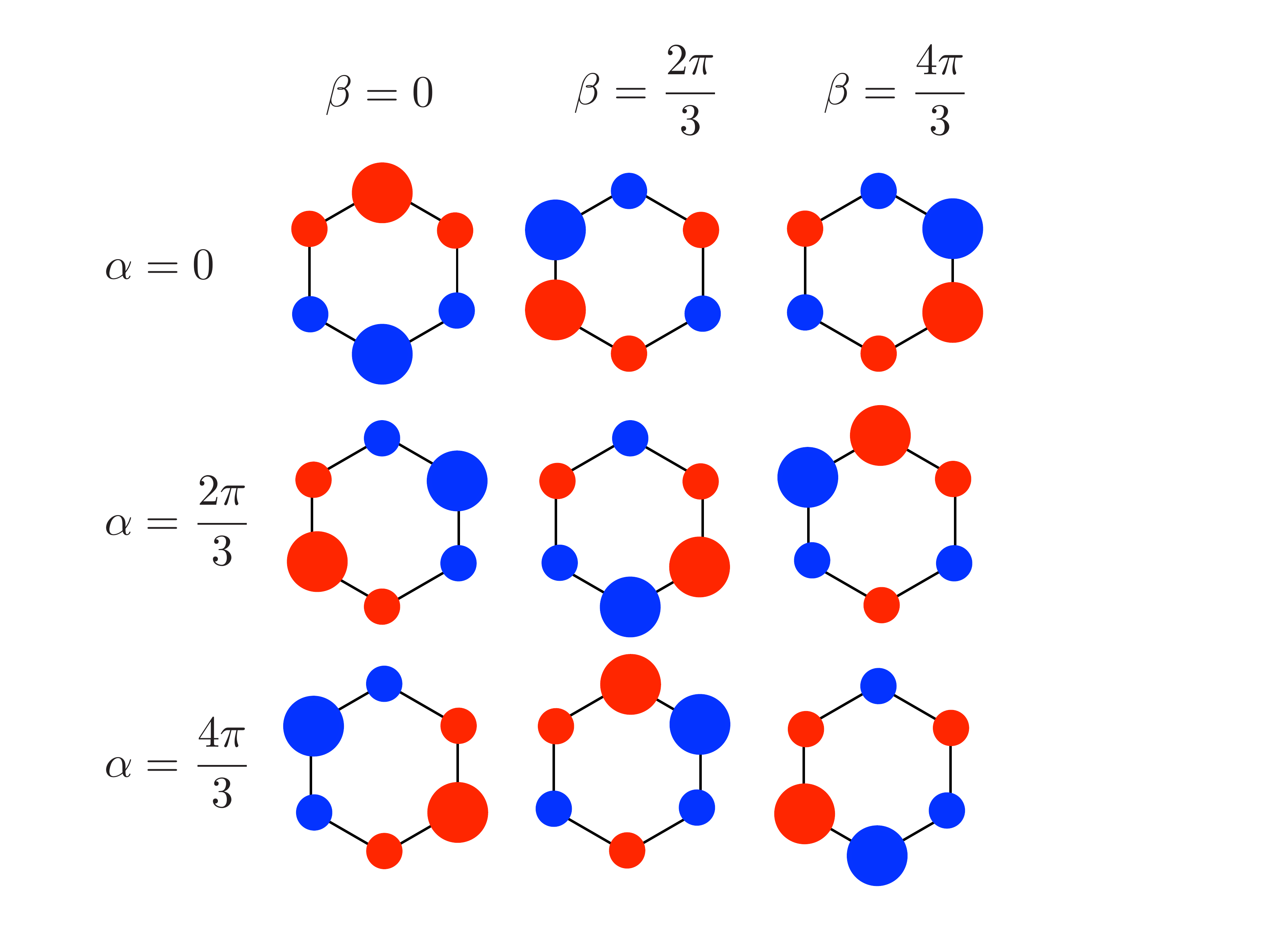}
 \caption{All 9 (18 with charge inversion) CDW$_3$ charge
   patterns. Charge modulation is shown relative to half-filling, with
   blue for positive modulation, and red for negative modulation. Different
   $\alpha$ correspond to $C_3$ rotations. Different $\beta$
   correspond to shifts of the unit cell.}
\label{figpatterns}
\end{figure}

The free energy is minimized for  $\v{A}^1\times\v{A}^2=0$, equivalent to $ \beta=\gamma$. Otherwise particle-hole symmetry would be broken, which is energetically unfavorable. 
Under this condition the lattice cubic and quartic terms terms are 
\begin{align}
L_{b_3}&=b_3A^3\sin(3\alpha)\sin(3\beta),\\
L_{b_4}&=b_4\phi A^3\cos(3\alpha)\cos(3\beta),
\end{align}
which are respectively minimized and maximized for $(\alpha,\beta) = \frac{2\pi}{3}(n,m)$ for integers $n,m=0,1,2$. 
Here $\alpha$ encodes translations and $\beta$ rotations, which together enumerate the 9 possible charge configurations, displayed in Fig.~\ref{figpatterns}. 
Naturally, the number of patterns is  doubled with charge inversion.  
Numerical simulations and semiclassical analysis~\cite{motruketalprb2015} have previously identified that the lattice ground state manifold is restricted to the same set of states.

The free energy can be partially minimized by letting $\Delta= x \rho$ and solving $\frac{\partial  f}{\partial \rho}=0$, which results in 
\begin{equation}
f_*(x) = -\frac{4 \pi ^2 \left(2  (x+2)^2 \sin \theta +  (x-1)^2 \cos \theta
   \right)^3}{(g_A)_*(g^2_\phi)_*27 (\pi -4)^2 (x-1)^4 (x+2)^2} \delta^3+\cdots.
  \end{equation}
In general, the higher order, non-universal terms $(\cdots)$ will select the minimizing value of $x$ or $\Delta$ (subject to constraints), for a given path $\theta$ through the critical surface. 
It is possible however to obtain the ground state  in the limiting cases by assuming generic bounding quartic terms.  
For $\theta \rightarrow \pi/2^-$ it is found that  $\Delta/\rho \rightarrow1^-$ and $ \phi\rightarrow0^+$. 
In the other limit, $\theta\rightarrow0^+$, it is found that $\Delta/\rho\rightarrow0^+$, $\phi\rightarrow A^+$.

The latter limit, $\theta\rightarrow0^+$, is precisely the condition for the broken symmetry state to host the previously discussed gapless semi-Dirac~\cite{dietletalprl2008,banerjeeetalprl2009,  yangetalprl2013} excitations, which disperse  quadratically in the direction defined by the polar angle $\beta$ and  linearly orthogonal to this.
 From the hybridization of down-folded Dirac valleys there is a condensation energy gain from the second set of bands that gap with  $\pm \abs{\phi+A}$. 
Such a metallic CDW$_3$ state with semi-Dirac quasiparticle excitations was predicted previously~\cite{christouetalprb2018} for the case of  a pure $V_2$ interaction. These findings were based on a self-consistent analysis of the free energy with the inclusion of RPA-type order parameter fluctuations. 
\begin{figure}[t]
 \includegraphics[width=0.4\textwidth]{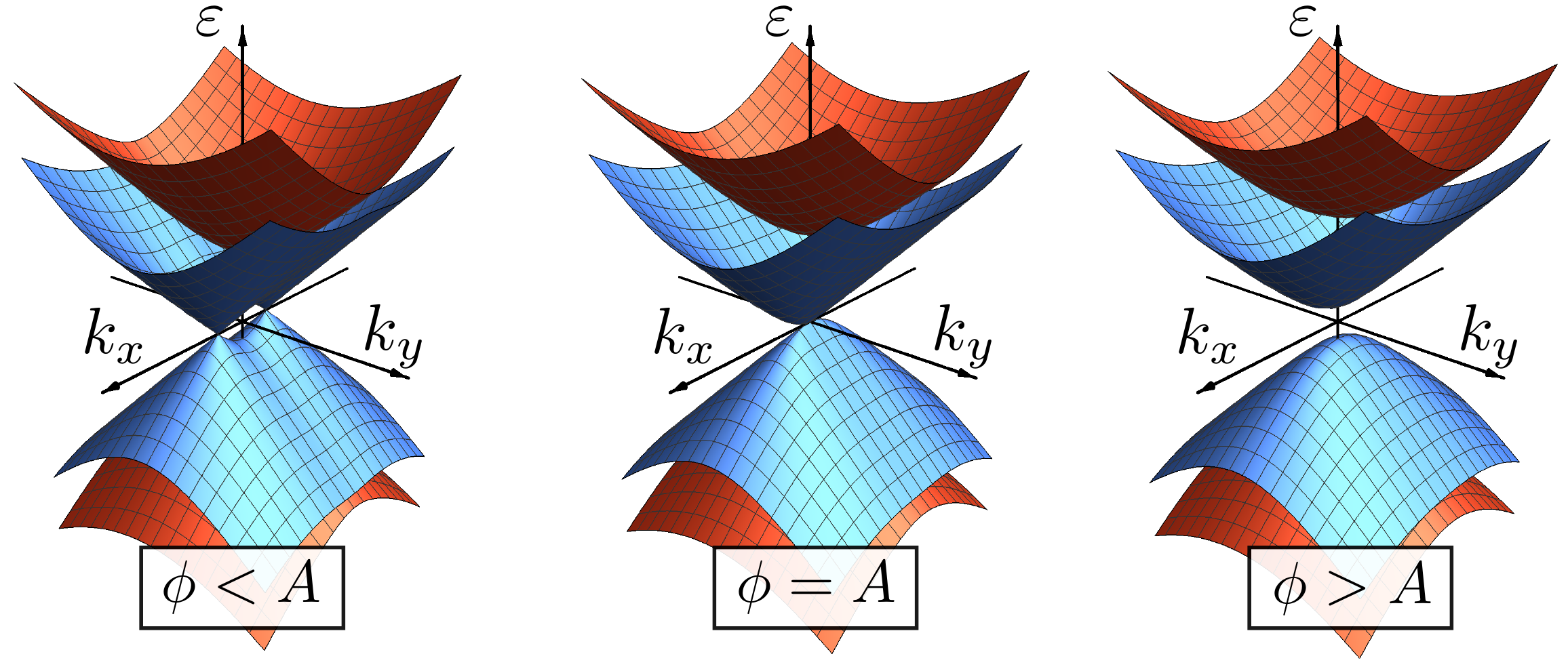}
 \caption{Topological phase transition between the Dirac semimetal
   ($\phi<A$) and band insulator $(\phi>A)$, tuned by complementary mass $\phi$ and 
 non-Abelian gauge $A$ fields. The topological critical point ($\phi=A$) hosts anisotropic excitations with orthogonal relativistic and non-relativistic directions.}
\label{topological}
\end{figure}

\subsection{Topological Phase Transitions}
The broken symmetry state with gapless semi-Dirac excitations $(\phi=A)$ can be interpreted as the critical point of the topological Lifshitz phase transition~\cite{yangetalprl2013} between a semimetal $(\phi<A)$ and a topologically trivial band insulator $(\phi>A)$, as shown in Fig.~\ref{topological}.
This notion for the $d=2$ topological critical point can be generalized to the full set of masses, 
$M^1$, $M^2$, $M^3$ \eqref{lmass} and non-Abelian gauge fields $\v{A}^1$, $\v{A}^2$, $\v{A}^3$ with $\v{A}^a=(A^a_x,A^a_y)$ \eqref{lgauge}.
This generalization accounts for Kekul\'e masses $(M^{1,2})$ and gauge fields $\v{A}^3$ that may be induced by strain on the honeycomb lattice \cite{vozmedianoetalphysicsreports2010}. 

Assuming full global $SU(2)$ gauge rotational symmetry the analytic part of the Landau free energy up to quartic order takes the form 
\begin{align}
f &= C_1 M^2 + C_2 A^2 + C_3 \epsilon^{abc} M^a (\v{A}^b\times \v{A}^c)_z +
    C_4 M^2 A^2 \notag \\ &\phantom{=}\, + C_5 M^a \v{A}^a \cdot M^b  \v{A}^b + C_6 M^4 + C_7
A^4  \notag \\ &\phantom{=}\, + C_8(\epsilon^{abc}\v{A}^b\times\v{A}^c)^2 + C_9
(\v{A}^a\cdot\v{A}^b) (\v{A}^a \cdot \v{A}^b),
\end{align}
where for this section the indices run over the full set $a,b,c = 1,2,3$ and where $C_i$ are unspecified coupling constants. 
The $C_3$ and $C_5$ terms must vanish to describe topological transitions. 
Subject to this constraint, and due to full SU(2) symmetry, the minimum of the free energy is now realized by any state obtained applying an SU(2) rotation to the $\phi=A$ state described above. For any such state, the Hamiltonian
\begin{equation}
\hat{H} = \v{k}\cdot\v{\alpha} +\alpha_z M^a T^a +\v{\alpha}\cdot \v{A}^a T^a,
\end{equation}
has a topological critical point at $\abs{M}=\abs{A}$. 
A finite $C_3$ term breaks particle-hole symmetry, whereas a finite $C_5$ term induces a band gap for all finite $M$ and $A$.

\section{Discussion}\label{secdiscussion}
We demonstrated that spontaneous lattice symmetry breaking in interacting Dirac fermion  systems is described by effective field theories in which dynamical order parameter fields minimally couple to Dirac fermions as a combination of mass and emergent gauge field components.
This is a departure from the common wisdom that mass channels solely dominate the energetic landscape~\cite{herbutetalprb2009, ryuetalprb2009}, which is an expectation motivated by the energetic gain upon condensing into an insulating state.

As a result, the ensuing criticality considered here is found to be beyond the GNY universality classes. 
The unconventional dynamical exponent $z>1$ indicates that the putative emergent Lorentz invariance associated with GNY criticality is violated. 
Our conclusions follow from a one-loop renormalization group analysis that is analytically controlled by the $\epsilon$-expansion, as well as the generalization to a large number $N$ of fermion flavors.

As a concrete example, the spontaneous lattice symmetry breaking due to CDW$_3$ order of Dirac fermions on the half-filled honeycomb lattice was analyzed. However, the conclusions regarding the multi-critical point are generic for the low-energy field theory. For a given starting microscopic model, the different order parameters will in general break different symmetries, which might be a combination of translations, point group or internal symmetries, but as long as the low energy theory is described by Dirac fermions in 2+1D, all order parameters will be represented by either mass fields or gauge fields and our theory can be used to describe the corresponding phase transitions.

The important role of emergent gauge fields has been recognized in the context of the Ising nematic transition  in $d$-wave superconductors~\cite{vojtaetalprl2000,vojtaetalijmpb2000,vojtaetalprb2000,kimetalprb2008,huhsachdevprb2008}, and there is indeed a close connection with the gauge sector of the field theory presented here.
In this case the fourfold lattice rotational symmetry is spontaneously reduced to twofold, and the effective Yukawa coupling in our notation reads $A_\text{nem} \Psi^\dagger[(\alpha_x+\alpha_y)+( \alpha_x - \alpha_y)T^3] \Psi$. 
However, dealing with fermion-induced criticality one might anticipate that the lattice terms are rendered irrelevant at the critical point. In this case, the problem should be
revisited, starting from a continuous rotational symmetry and the full set of non-Abelian gauge fields. 

Similarly, on the $\pi$-flux lattice~\cite{weeksfranzprb2010,jiaetalprb2013}, sublattice $A_\text{sub}\Psi^\dagger \alpha_x T^1\Psi$ and stripe $A_\text{str}\Psi^\dagger \alpha_x T^2\Psi$ charge order parameters both couple as emergent gauge fields, but have thus far not been analyzed as such. 
This case is interesting because interaction induced Haldane-like quantum Hall order couples as a mass field, but is destabilized by beyond mean-field fluctuations, much like the CDW$_3$ case. 
In both instances, the existence of emergent gauge fields is tied to the breaking of lattice symmetries. 

That $z>1$ raises the possibility that  long-range Coulomb  interactions \cite{herbutprl2006} 
\begin{equation}
\hat{n}(\v{r})\frac{e^2}{\abs{\v{r}-\v{r}^\prime}}\hat{n}(\v{r^\prime})\rightarrow -i \Psi^\dagger A^0_0 \Psi +\frac{1}{2e^2} A^0_0 \abs{\v{q}} A^0_0,
\end{equation}
could be relevant and provide further non-trivial scaling at this novel critical point~\cite{isobeprl2017}.
The one-loop RG equation for the Coulomb coupling~\cite{herbutetalprb2009}
\begin{equation}
\dl e^2 \approx (z-1)e^2 - \delta_{d,3}e^4,
\end{equation}
demonstrates that $z>1$ defines $e$ as a relevant perturbation. At the new CDW$_3$ fixed point it is expected that $z\sim 1 + (g_A^2)_*-x e^2$  $(x>0)$, suggesting such relevance.
In contrast, the Coulomb interaction is irrelevant at the $d=2$ GNY fixed point $z =1 - x e^2$ $(x>0)$~\cite{herbutprl2006,herbutetalprb2009,royetaljhep2016,janssenheprb2017}.

Recently, there has been considerable interest in the properties of topological quantum critical points~\cite{yangetalnatphys2014,isobeprl2017,uchoaseoprb2017,royfosterprx2018,linketalprl2018,lietalprb2018,hanetalprb2018,hanetalprl2019,surroyprl2019,uryszeketalarxiv2019}, which in $d=2$ are commonly described by effective Hamiltonians of the semi-Dirac form $\hat{H} = k_x^2 \sigma_x + k_y\sigma_y$. 
Here anisotropic velocity renormalization needs to be  regulated with non-perturbative infrared loop resummations~\cite{surroyprl2019,uryszeketalarxiv2019}.

Our work shows that complementary combinations mass and non-Abelian gauge fields provide  a natural playground for the study of topological quantum phase transitions. 
Lifshitz transitions of merging  Dirac cones are observed when tuning through the broken-symmetry states close to the multi-critical point. 
These insights could be relevant for a range of systems, including  black phosphorus~\cite{kimetalscience15, kimetalprl2017}, optical honeycomb lattices~\cite{tarruelletalnature2012,uehlingeretalprl2013,tarnowskiprl2017},  artificial graphene~\cite{gomesetalnature2012}, TiO$_2$/VO$_2$ interfaces~\cite{pardopickettprl2009,banerjeeetalprl2009},  and $\alpha$(BEDT-TTF)$_2$I$_3$~\cite{kobayashietaljpsj2007}.

A related open question is whether there is a similar description of other exotic Lorentz violating fermions, such as the multifolds in topological chiral crystals~\cite{bradlynetalscience2016,sanchezetalnature2019,schroteretalnatphys2019}, and what this means for their quantum critical properties~\cite{royetalprl2018,boettcherarxiv2019}.

{\it Acknowledgements.}$-$ We thank  Andrey
Chubukov,  Felix Flicker, Andrew Green,  Andrew James,  Vladimir
Juri\vaccent{c}i\'{c},  Andriy Nevidomskyy,  Siddharth Parameswaran, Bitan Roy, Bruno
Uchoa and Jasper van Wezel for useful discussions. F.~K. acknowledges financial support from EPSRC under Grant EP/P013449/1.

\appendix
\section{Derivation of the Effective Field Theory}
\label{appendixderivation}
Here we obtain the effective field theory from the extended Hubbard model of spinless fermions on the honeycomb lattice.
We accommodate for broken translational symmetry
with a 6-site enlarged unit cell that covers a honeycomb plaquette. In
this case, the
lattice vectors are
\begin{equation}
\v{a}_{1}=\frac{3a}{2}(\sqrt{3},1),\,\,
\v{a}_{2}=\frac{3a}{2}(-\sqrt{3},1), \,\,
\v{a}_{3}=- \v{a}_1 - \v{a}_2,
\end{equation}
and the basis is
\begin{equation}
\v{c}=(c_{A_1}, c_{A_2}, c_{A_3}, c_{B_1}, c_{B_2}, c_{B_3}). \label{basis}
\end{equation}
The corresponding reciprocal lattice vectors of the down-folded Brillouin zone are
$\v{b}_{1,2}=\frac{2\pi}{3\sqrt{3}a}(\pm1,\sqrt{3})$.

\noindent The non-interacting Hamiltonian is 
\begin{align}
H_t & = \sum_{\v{k}}\v{c}^{\dagger}(\v{k})H_t(\v{k})\v{c}(\v{k}),\\
H_t(\v{k})&=
\begin{pmatrix}0&T^\dagger(\v{k})\\T(\v{k})&0\end{pmatrix},\\
T(\v{k})&=-t\begin{pmatrix}
1&1&e^{i \v{k}\cdot\v{a}_2}\\
1&e^{i \v{k}\cdot\v{a}_3}&1\\
e^{i \v{k}\cdot\v{a}_1}&1&1
\end{pmatrix}.
\end{align}
Due to down-folding, both Dirac points reside at the $\Gamma$ point
($\v{k}=\v{0}$). We integrate out high energy modes by projecting into
the low-energy Dirac subspace $\Psi = P \v{c}$, and obtain
the non-interacting Dirac Hamiltonian
\begin{equation}
\mathcal{H}_t(\v{p}) = P\left\{ H_t(\v{0}) +
                       [\nabla_{\v{k}}H_t(\v{k})]_{\v{k}=\v{0}}\cdot\v{p}
                       \right\}P^\dagger  =v_F \v{p}\cdot\v{\alpha}.
\end{equation}
The projection $P$ is formulated from the low
energy (row) eigenvectors of $H_t(\v{k}=\v{0})$,
\begin{equation}
P_{0}= \frac{1}{\sqrt{6}}\begin{pmatrix}
-1&-1&2&0&0&0\\
\sqrt{3}&-\sqrt{3}&0&0&0&0\\
0&0&0&-\sqrt{3}&\sqrt{3}&0\\
0&0&0&-1&-1&2
\end{pmatrix}.
\end{equation}
For convenience we apply the additional unitary transformation, such that the projection
is 
\begin{equation}
  P=
  e^{- i\frac{\pi}{4} \sigma_z\otimes\tau_z}
  e^{- i \frac{2\pi}{3} \sigma_0\otimes\tau_z}
  e^{- i \frac{\pi}{4} \sigma_0\otimes\tau_x}P_0.
  \label{projection}
\end{equation}
The first exponential transforms into the basis $\v{\alpha}$. The second enacts a
coordinate transformation. The third translates the definition of the unit cell.

Now we can obtain the low-energy theory of the charge ordering. First
we decouple the interaction in the charge channel with the
Hubbard-Stratonovich transformation \eqref{HS} where the explicit for of the 
coupling matrix is

\begin{widetext}
\begin{equation}
  \v{V}_{\v{k}}=-\frac{1}{2}
  \begin{pmatrix}
    0&V_2e_{\bar{1}3}&V_2e_{\bar{1}2}&V_1&V_1&V_1 e^{i \v{k}\cdot\bar{\v{a}}_1}\\
    V_2e_{1\bar{3}}&0&V_2e_{2\bar{3}}&V_1&V_1 e^{i \v{k}\cdot\bar{\v{a}}_3}&V_1\\
    V_2e_{1\bar{2}}&V_2e_{\bar{2}3}&0&V_1 e^{i \v{k}\cdot\bar{\v{a}}_2}&V_1&V_1\\
    V_1&V_1&V_1 e^{i \v{k}\cdot\v{a}_2}&0&V_2e_{2\bar{3}}&V_2e_{\bar{1}2}\\
    V_1&V_1 e^{i \v{k}\cdot\v{a}_3}&V_1&V_2e_{\bar{2}3}&0&V_2e_{\bar{1}3}\\
    V_1 e^{i \v{k}\cdot\v{a}_1}&V_1&V_1&V_2e_{1\bar{2}}&V_2e_{1\bar{3}}&0
  \end{pmatrix},
  \label{vmatrix}
\end{equation}
with $  e_{nm}=1+e^{i\v{k}\cdot \v{a}_n}+e^{i\v{k}\cdot \v{a}_m}$
and $\v{a}_{\bar{n}}=-\v{a}_n$, $n=1,2,3$.
Applying the projection, we obtain the local interaction term
\begin{align}
L_{V}& = \Psi^\dagger\{
(V_2 -
  V_1/2)[\rho_{A_1}+\rho_{A_2}+\rho_{A_3}-\rho_{B_1}-\rho_{B_2}-\rho_{B_3}
                         ]\alpha_3T^3
                         +\frac{V_2}{4}[\rho_{A_1}-2\rho_{A_2}+\rho_{A_3}-\rho_{B_1}+2\rho_{B_2}-\rho_{B_3}
  ]\alpha_xT^1\notag\\
&\phantom{=}\,\,+\frac{\sqrt{3}V_2}{4}[-\rho_{A_1}+\rho_{A_3}-\rho_{B_1}+\rho_{B_3}
                         ]\alpha_yT^1
                         +\frac{\sqrt{3}V_2}{4}[-\rho_{A_1}+\rho_{A_3}+\rho_{B_1}-\rho_{B_3}
  ]\alpha_xT^2\notag\\
&\phantom{=}\,+\frac{V_2}{4}[-\rho_{A_1}+2\rho_{A_2}-\rho_{A_3}-\rho_{B_1}+2\rho_{B_2}-\rho_{B_3}
  ]\alpha_yT^2  +(V_2 +
  V_1/2)[\rho_{A_1}+\rho_{A_2}+\rho_{A_3}+\rho_{B_1}+\rho_{B_2}+\rho_{B_3}]
                           \} \Psi \notag\\
  &\phantom{=}\,+ V_1
(\rho_{A_1}+\rho_{A_2}+\rho_{A_3})(\rho_{B_1}+\rho_{B_2}+\rho_{B_3})+ 3 V_2 (
                \rho_{A_1}\rho_{A_2}+\rho_{A_2}\rho_{A_3}+\rho_{A_3}\rho_{A_1}+\rho_{B_1}\rho_{B_2}+\rho_{B_2}\rho_{B_3}+\rho_{B_3}\rho_{B_1}).
\end{align}
\end{widetext}
Here the gradient terms have been suppressed in the interest of
brevity. The relevant gradient terms are systematically included
by the one-loop fermion bubbles. 

\section{RG Equations in $d=2$}\label{appendixrgequations}
The RG equations are calculated directly in the physical
dimensions $d=2$, with the expansion controlled by large $N$,
\begin{equation}
	\Psi^\dagger\Psi \rightarrow
       \sum_{n=1}^N\Psi^\dagger_n\Psi_n,\,
      (g^2_{\phi,A}, \lambda_i) \rightarrow \frac{8\pi \Lambda}{N}(g^2_{\phi,A},\lambda_i).
    \end{equation}
The RG equations for the velocities are
  \begin{align}
\dl v_F &= v_F \bigg[z-1+\frac{1}{N} \bigg(\frac{g_{\phi }^2
   \left(c_{\phi }-2 v_F\right)}{c_{\phi } v_F
          \left(c_{\phi }+v_F\right){}^2}\notag \\
          &\phantom{=}\, -\frac{8 g_A^2}{c_A
   \left(c_A+v_F\right){}^2}\bigg)\bigg],\\
 \dl c_\phi^2 &=2 (z-1) c_{\phi }^2+\frac{g_{\phi }^2 \left(v_F^2-2 c_{\phi
   }^2\right)}{v_F^3}, \\
  \dl  c_A^2&=2
   (z-1) c_A^2+\frac{g_A^2 \left(v_F^2-2 c_A^2\right)}{2 v_F^3}.
  \end{align}
   For the Yukawa couplings we obtain
   \begin{align}
  \dl g_\phi^2 &=g_{\phi }^2 \bigg[3 z-2-\frac{2
   g_{\phi }^2}{v_F^3}+\frac{1}{N} \bigg(-\frac{16 g_A^2
   \left(c_A+2 v_F\right)}{c_A v_F
   \left(c_A+v_F\right){}^2}\notag \\  &\phantom{=}\, -\frac{4 g_{\phi }^2
   \left(c_{\phi }+2 v_F\right)}{c_{\phi } v_F
   \left(c_{\phi }+v_F\right){}^2}\bigg)\bigg] ,\\
\dl g_A^2  &=g_A^2 \bigg[3
   z-2-\frac{g_A^2}{v_F^3}+\frac{1}{N} \bigg(-\frac{16 g_A^2}{c_A
   \left(c_A+v_F\right){}^2} \notag \\  &\phantom{=}\,-\frac{2 g_{\phi }^2
   \left(c_{\phi }+2 v_F\right)}{c_{\phi } v_F
   \left(c_{\phi
   }+v_F\right){}^2}\bigg)\bigg] .
   \end{align}
   The RG equations for the order parameter self-interactions are
   \begin{align}
 \dl \lambda_\phi  &=\lambda _{\phi } \left(3 z-2-\frac{4 g_{\phi
   }^2}{v_F^3}\right)+\frac{g_{\phi }^4}{v_F^3}\notag \\ & \phantom{=}\,+\frac{1}{N} \left(-\frac{4 \lambda _{\phi 
   A}^2}{c_A^3}-\frac{36 \lambda _{\phi }^2}{c_{\phi
   }^3}\right) ,\\
 \dl \lambda_A &=\lambda _A
   \left(3
   z-2 -\frac{2 g_A^2}{v_F^3}\right)-\frac{g_A^4}{8 v_F^3} \notag \\ & \phantom{=}\,+\frac{1}{N} \left(-\frac{\lambda _{\phi  A}^2}{c_{\phi
   }^3}-\frac{48 \lambda _A^2}{c_A^3}-\frac{4 \lambda
   _A \lambda _{\text{YM}}}{c_A^3}-\frac{\lambda
   _{\text{YM}}^2}{c_A^3}\right) ,\\
  \dl \lambda_{\text{YM}}&=\lambda _{\text{YM}}
   \left(3 z-2 -\frac{2 g_A^2}{v_F^3}\right)+\frac{3
   g_A^4}{2 v_F^3} \notag \\ & \phantom{=}\,+\frac{1}{N} \left(-\frac{8 \lambda
   _{\text{YM}}^2}{c_A^3}-\frac{48 \lambda _A \lambda
   _{\text{YM}}}{c_A^3}\right) , \\
   \dl \lambda_{\phi A}&=\lambda _{\phi  A}
   \left(3
   z-2 -\frac{g_A^2+2 g_{\phi }^2}{v_F^3}\right)+\frac{3 g_A^2 g_{\phi }^2}{v_F^3} \notag \\ & \phantom{=}\,+\frac{1}{N} \bigg(\frac{32 \lambda _{\phi  A}^2}{c_A
   c_{\phi } \left(c_A+c_{\phi }\right)}-\frac{24
   \lambda _A \lambda _{\phi  A}}{c_A^3}\notag \\ & \phantom{=+\frac{1}{N}\bigg(}\,-\frac{12
   \lambda _{\phi } \lambda _{\phi  A}}{c_{\phi
   }^3}-\frac{2 \lambda _{\text{YM}} \lambda _{\phi 
   A}}{c_A^3}\bigg) .
  \end{align}
  The RG equations for the order parameter masses are
  \begin{align}
   \dl m^2_\phi&=2 m_{\phi }^2
   \left(z-\frac{g_{\phi }^2}{v_F^3}\right) +\frac{1}{N} \bigg(-\frac{12 \lambda _{\phi }
   m_{\phi }^2}{c_{\phi }^3}\notag \\ & \phantom{=}\,-\frac{8 m_A^2 \lambda
   _{\phi  A}}{c_A^3}+\frac{8 \Lambda ^2 \left(3 c_A
   \lambda _{\phi }+2 c_{\phi } \lambda _{\phi 
   A}\right)}{c_A c_{\phi }}\bigg) , \\
    \dl m^2_A &= m_A^2 \left(2
   z-\frac{g_A^2}{v_F^3}\right)+ \frac{1}{N} \bigg(-\frac{2 m_{\phi }^2 \lambda _{\phi 
   A}}{c_{\phi }^3}\notag \\ & \phantom{=}\,-\frac{2 m_A^2 \left(12 \lambda
   _A+\lambda _{\text{YM}}\right)}{c_A^3}\notag \\ & \phantom{=}\,+\frac{4
   \Lambda ^2 \left(c_A \lambda _{\phi  A}+c_{\phi }
   \left(12 \lambda _A+\lambda
   _{\text{YM}}\right)\right)}{c_A c_{\phi
   }}\bigg).
  \end{align}
  The lattice allowed cubic term is irrelevant in this treatment, as can be seen from the RG equation
  \begin{equation}
\dl b_3^2 = b_3^2\left[5 z-2 -\frac{3 g_A^2}{v_F^3}-\frac{12 \frac{1}{N} \left(4 \lambda _A+\lambda
   _{\text{YM}}\right)}{c_A^3}\right] + \mathcal{O}(b_3^4),
\end{equation}
with the dimensionless coupling $b_3^2\rightarrow \frac{8\pi \Lambda}{N}b_3^2$. 
At the CDW$_3$ fixed point this reduces to 
\begin{equation}
\dl b_3^2 = 2 b_3^2\frac{-57+32\sqrt{2}}{N}  + \mathcal{O}(b_3^4).
\end{equation}
Therefore, to this order $b_3$ is an irrelevant perturbation and the fixed point value is $(b_3)_*=0$. In $N\rightarrow \infty$, $b_3$ is marginal to all loop orders following the argument in Ref.~\cite{schererherbutprb2016}. A similar RG equation is obtained for $\tilde{b}_3$.

\end{document}